\documentclass[twocolumn,aps,pra,floatfix,showpacs]{revtex4}

\usepackage{graphicx,color}
\usepackage{dcolumn,bm}
\usepackage{mathbbol}
\usepackage{amsmath,amssymb}
\usepackage[latin9]{inputenc} 
\usepackage[sans]{dsfont}
\usepackage[mathscr]{euscript}

\newtheorem{Prop}{Proposition}
\newtheorem{defin}{Definition}
\newtheorem{thm}{Theorem}
\newtheorem{cor}{Corollary}
\newtheorem{lemma}{Lemma}

\newcommand{\ket}[1]{|#1\rangle}
\newcommand{\bra}[1]{\langle #1|}

\newcommand{\Hi}{\mathcal{H}}

\newcommand{\supp}{\rm{supp}}

\newcommand{\beq}{\begin{equation}}
\newcommand{\eeq}{\end{equation}}
\newcommand{\beqa}{\begin{eqnarray}}
\newcommand{\eeqa}{\end{eqnarray}}
\newcommand{\beqan}{\begin{eqnarray*}}
\newcommand{\eeqan}{\end{eqnarray*}}

\newcommand{\C}{\mathbb{C}}

\newcommand{\R}{\mathbb{R}}
\newcommand{\qed}{\hfill $\Box$ \vskip 2ex}

\newcommand{\range}{\rm{range}}

\renewcommand{\ker}{\rm{ker}}
\renewcommand{\span}{\rm{span}}
\newcommand{\rank}{\rm{rank}}

\newcommand{\tr}{\rm{Tr}}

\begin{document}

%\setlength{\textheight}{8.0truein}    %FOR 2ND PAGE ONWARDS

%%\runninghead{Steady-state entanglement by engineered 
%            quasi-local Markovian dissipation}
%            {F. Ticozzi and L. Viola}

%\normalsize\textlineskip
%\thispagestyle{empty}
%\setcounter{page}{1}

%\copyrightheading{Vol.}{No.}{Year}{Page Nos.}
%\copyrightheading{0}{0}{2003}{000--000}

%\vspace*{0.88truein}

%\alphfootnote
%
%\fpage{1}

\title{STEADY-STATE ENTANGLEMENT BY\\ ENGINEERED QUASI-LOCAL MARKOVIAN DISSIPATION}

\author{Francesco Ticozzi}
\email{ticozzi@dei.unipd.it}
\affiliation{Dipartimento di Ingegneria dell'Informazione,
Universit\`a di Padova, via Gradenigo 6/B, 35131 Padova, Italy}\affiliation{\mbox{Department of Physics and Astronomy,
Dartmouth College, 6127 Wilder Laboratory, Hanover, NH 03755, USA}}
\author{Lorenza Viola}
\email{lorenza.viola@dartmouth.edu}
\affiliation{\mbox{Department of Physics and Astronomy,
Dartmouth College, 6127 Wilder Laboratory, Hanover, NH 03755, USA}}

\date{\today}

\begin{abstract}
We characterize and construct time-independent Markovian dynamics 
that drive a finite-dimensional multipartite quantum system into a target (pure) 
entangled steady state, subject to physical locality constraints. In situations where the 
desired stabilization 
task can not be attained solely based on local dissipative means, we allow for 
local Hamiltonian control or, if the latter is not an option, we suitably restrict 
the set of admissible initial states.  In both cases, we provide 
algorithms for constructing a master equation 
that achieves the intended objective and show how this can genuinely extend the 
manifold of stabilizable states.
In particular, we present quasi-local control protocols for dissipatively 
engineering multipartite GHZ ``cat'' states and W states on $n$ qubits. 
For GHZ states, we find that no scalable procedure exists for achieving stabilization 
from arbitrary initial states, whereas this is possible for a target W state by a suitable 
combination of a two-body Hamiltonian and dissipators. Interestingly, 
for both entanglement classes, we show that quasi-local stabilization may be scalably 
achieved conditional to initialization of the system in a large, appropriately chosen 
subspace.\end{abstract}

\maketitle

\noindent {\em Keywords:} Quantum control, engineered dissipation, entanglement,
quantum dynamical semigroups.

\section{Introduction}
Generating and manipulating highly non-classical entangled states in a robust 
scalable fashion is a central goal across Quantum Information Processing (QIP) 
\cite{nielsen-chuang} and quantum engineering \cite{milburn}.   While traditional 
circuit-based schemes are based on the application of suitable coherent control, 
implementing a desired sequence of quantum gates on a known fiducial state, 
progress in controlling open quantum systems is prompting a systematic 
exploration of incoherent control schemes using {\em engineered dissipation}. 
Access to dissipative control parameters, alone or in conjunction with Hamiltonian
ones, may generally allow for enhanced efficiency and added flexibility in achieving 
relevant QIP tasks.  Notably, quantum simulation of arbitrary open-system dynamics 
on a quantum computer may be accomplished by suitable dissipative control on just 
one more ``ancillary'' qubit than needed to simulate closed-system dynamics 
\cite{LloydViola}, as demonstrated in recent trapped-ion experiments 
\cite{barreiro,schindler} (see also \cite{Thomas}).  
Likewise, dissipative quantum-state preparation 
protocols may allow for arbitrary initial states to be driven into a desired target 
state by ``all-to-one'' control laws that have no counterpart for unitary evolutions, 
with the potential for inherent robustness against transient perturbations. As a 
result, schemes for dissipative quantum control and generation of entangled states 
of light and matter via ``reservoir engineering'' have been extensively investigated 
theoretically in different contexts, see e.g. 
\cite{davidovich,poyatos,noon,diehl,Kraus-entangled,ticozzi-QDS,ticozzi-markovian,Verstraete2009,Sophie,ticozzi-mixed,Almut,Pechen,diehlTop,Chen,Dorner,Lukin,NaokiTrans,Naoki2012,aah}.

As control methodologies that rely on engineered dissipation become 
increasingly attractive experimentally \cite{barreiro,polzik-entanglement,polzik2, devoret}, 
it is essential to develop rigorous control-theoretic characterizations of the set of 
target states achievable in open quantum systems given realistic control 
resources, and to devise constructive control design protocols.
In this context, an important 
task is provided by dissipative entanglement engineering in multipartite systems 
evolving according to a continuous-time quantum Markov process \cite{Alicki}.  
It is worth stressing that designing Markovian dynamics that admits a desired 
(pure or mixed) target state as its unique steady state (that is, achieves 
global asymptotic stability at the target) is relatively straightforward in the 
absence of control constraints, and different constructive strategies exist. In 
fact, application of a single time-independent Hamiltonian and Markovian noise 
channel suffices in principle in the generic case  \cite{ticozzi-mixed}. Alternatively, 
if the system is completely controllable in the absence of dissipation and the control 
actions can be enacted fast with respect to the noise time-scales, it clearly 
suffices to identify a purely dissipative dynamics that stabilizes any state with the 
same spectrum as the target, followed by rapid unitary control in order to suitably 
rotate such state into the target \cite{Pechen}.  

In practice, however, physical evolutions are not described by arbitrary Markovian
master equations, and available coherent and incoherent controls are inevitably 
constrained.  In particular, physically admissible Hamiltonian and noise (Lindblad) 
operators are typically described by operators that act non-trivially on finite subsets of 
subsystems, that is, are {\em Quasi-Local} (QL) relative to the given tensor-product 
decomposition. Necessary and sufficient conditions for a pure entangled state to 
be stabilizable in the absence of pre-existing ``drift'' dynamics under QL constraints 
have been obtained in \cite{ticozzi-ql}, assuming {\em purely dissipative means} 
(so-called Dissipative QL Stabilization, DQLS for short). 

In this work, we continue and substantially expand our investigation, with the 
twofold goal of (i) obtaining a complete characterization of 
the DQLS setting in the presence of non-trivial drift dynamics 
\cite{bertinoro}; (ii) defining and
analyzing alternative QL settings, which may allow for a desired
stabilization task to be achieved in cases where purely dissipative
control is insufficient.  While we shall in general allow for both
{\em time-independent dissipative and Hamiltonian} (coherent) control
resources, no access to a universal set of tunable Hamiltonians 
%FT: com'era poteva confondere con coherent feedback, e con il fatto che i nostri Hams sono abbastanza da generare la Lie Algebra -- il punto e` che non sono time-dependent
%% LV: Concordo.  Slight reword 
nor auxiliary
quantum resources will be assumed, preventing switching schemes along
the lines envisioned in \cite{Pechen} or \cite{Naoki2012} from being
viable.  Specifically, two natural complementary strategies will be
pursued: to either augment the available control resources, by
suitably combining (time-independent) QL Hamiltonian control with
dissipation; or, if Hamiltonian control is not an option or is
otherwise unfeasible, to insist on purely dissipative stabilization by
restricting the initial condition to an appropriately chosen ``attraction basin''.

The content is organized as follows. We begin in Section \ref{QLmme}
by introducing the relevant class of dynamical models and different QL
stabilization settings, along with required known results on DQLS
states.  Hamiltonian-assisted and conditional QL stabilization are
first analyzed in Section \ref{sec:qls} and Section
\ref{sec:conditional}, respectively, under the simplifying assumption
that internal drift dynamics is absent or can be ignored.  In both
cases, we develop necessary conditions that the control Hamiltonian
or, respectively, the initialization subspace must obey in order for
the target state to be QL stabilizable, and then present a {\em
constructive randomized algorithm} that outputs a choice of
stabilizing controls with unit probability.  We reconsider and address
the role of underlying Markovian drift dynamics in Section
\ref{sec:drift}, identifying necessary conditions for the desired
stabilization to be compatible with the drift, and showing how
different scenarios may then be tackled by adapting our previous
results.  Section \ref{sec:example} present explicit QL stabilization
protocols for dissipative engineering of GHZ and W states in $n$-qubit
systems.  While none of these highly
entangled states can be generated solely based on QL dissipative means, except for trivial exceptions 
\cite{ticozzi-ql}, we show that for both classes QL stabilization is
achievable in a {\em scalable} fashion for a large subspace of initial
states. For GHZ states, we additionally establish that no scalable
procedure exists for achieving stabilization from arbitrary initial
states, whereas arbitrary initial states may be driven to a target W
state by a suitable combination of two-body Hamiltonian and
dissipators. Concluding remarks are given in Section \ref{sec:end}. 
We include in separate Appendices the proof of convergence of our 
randomized construction as well as, for completeness and direct reference,  an 
algorithm for verifying global stability, originally developed in \cite{ticozzi-NV}, 
that is required in the proof. 

%\vspace*{1pt}\textlineskip 
%--------------------------------------------------------------------------------------------
\section{Preliminaries and background}

\subsection{Quasi-local quantum dynamical semigroups}
\label{QLmme}

We consider a finite-dimensional multipartite open quantum system $S$
consisting of $n$ (distinguishable) subsystems, defined on a tensor-product 
Hilbert space
$$\Hi=\bigotimes_{a=1}^n\Hi_a, \;\;\; \;a=1,\ldots,n,\;\; 
\text{dim}(\Hi_a)=d_a, \, \text{dim}(\Hi)=d. $$
\noindent 
Let ${\mathfrak D}(\Hi)$ denote the (convex) set of physical states
(density operators) of $S$, that is, trace-one, positive semi-definite
operators with support on $\Hi$.  In analogy with probability
distributions, we shall denote the support of $\rho \in {\mathfrak
D}(\Hi)$ by $\supp(\rho)=\range(\rho)$. It is then natural to use ${\mathfrak D}(\Hi')$ 
to describe the set of density operators with
support contained in a subspace $\Hi' \subseteq \Hi$. 

In many QIP scenarios of interest, $S$ undergoes continuous-time
dissipative dynamics due to the coupling to a (physical or engineered)
Markovian environment, in which case the resulting dynamics is
described by a {\em Lindblad master equation}
\cite{Alicki,nielsen-chuang}. In units where $\hbar=1$, we may write
\beqa
\label{eq:lindblad}
\dot\rho(t)&\equiv& {\cal L}[\rho(t)] \\&=&-i[H,\,\rho(t)]+\sum_k \Big(
L_k\rho(t)L^\dag_k-\frac{1}{2}\{L_k^\dag L_k,\,\rho(t)\} \Big), \nonumber 
\eeqa
\noindent 
in which ${\cal L}$ represents the most general form for the generator of a
completely-positive, trace-preserving Markov semigroup, $\{ {\cal T}_t
\equiv e^{{\cal L}t},t\geq 0\}$, acting on ${\mathfrak D}(\Hi)$. Here,
$H$ is an Hermitian operator associated with the Hamiltonian of $S$,
whereas the Lindblad (or noise) operators $\{L_k\}$ specify the
non-Hamiltonian component of the generator, resulting in non-unitary
irreversible dynamics.  It is worth recalling that, for a given
Markovian generator, the decomposition into Hamiltonian and
dissipative part in Eq. \eqref{eq:lindblad} is {\em not} unique, in
the sense that 
$${\cal L} (H, \{ L_k \}) \equiv {\cal L} (\tilde{H}, \{ \tilde{L}_k
\}),$$
\noindent 
provided that the relevant operators are redefined as follows, see
{\em e.g.}  Lemma 2 in \cite{ticozzi-QDS}:
\begin{equation}
\label{eq:equiv}
\tilde{L}_k \equiv L_k + c_k I , \;\;\;\; \tilde{H}\equiv H +
\frac{i}{2} \sum_k (c_k^\ast L_k -c_k L^\dagger_k), \;\; c_k \in
{\mathbb C}.
\end{equation}
%% LV: Checked that sign is correct

Given a dynamical evolution of the above form, locality constraints
are imposed by requiring both the Hamiltonian and each Lindblad
operator to have a non-trivial (non-identity) action only on certain
subsets of subsystems, which may be distinguished by the geometry of
the system and/or a physical coupling topology, and which we shall
henceforth refer to as {\em neighborhoods}. Following
\cite{ticozzi-ql}, neighborhoods $\{ {\cal N}_j \}$ can be specified
in full generality as subsets of the set of indexes labeling the
subsystems, that is,
\[{\cal N}_j\subsetneq\{1,\ldots,n\}, \quad j=1,\ldots, M.\]
\noindent 
Once a neighborhood structure is assigned on $\Hi$, a list of {\em
reduced neighborhood states} $\{ \rho_{{\cal N}_j} \}$ can also be
naturally associated to every state of $S$:
\begin{equation} 
\label{redstate} 
\rho_{{\cal N}_j} \equiv \mbox{Tr}_{\bar{\cal N}_j}(\rho),
\quad \rho \in {\mathfrak D}(\Hi), \;\; j=1,\ldots, M,
\end{equation}
where $\tr_{\bar{\cal N}_j}$ indicates the partial trace over the {
tensor complement} of the neighborhood ${\cal N}_j,$ namely,
$\Hi_{\bar{\cal N}_j}=\bigotimes_{a\notin{\cal N}_j}\Hi_a$.  A
Lindblad master equation as in Eq. (\ref{eq:lindblad}) will be called
QL if {\em both} its Hamiltonian and each of the noise operators are
QL, according to the following:

\begin{defin}
A {\em Lindblad operator $L$ is QL} if there exists a neighborhood
${\cal N}_j$ such that:
\[L \equiv L_{{\cal N}_j}\otimes I_{\bar{\cal N}_j},\] 
where $L_{{\cal N}_j}$ accounts for the action of $L$ on the
subsystems in ${\cal N}_j$, and $I_{\bar{\cal
N}_j}:=\bigotimes_{a\notin{\cal N}_j}I_a$ is the identity on the
remaining ones.  Similarly, a {\em Hamiltonian $H$ is QL} if it admits
a decomposition into a sum of QL terms:
\[H = \sum_j H_j,
\quad H_j \equiv H_{{\cal N}_j}\otimes I_{\bar{\cal N}_j}.\]
\noindent 
\end{defin} 

It is immediate to verify that the QL property is well defined with
respect to the freedom in the representation of the Markov generator,
since the transformation in Eq. (\ref{eq:equiv}) preserves the 
quasi-locality character of $H$ and $\{L_k\}$.  
%% LV: Concordo col tuo punto, anzi ho cambiato does not affect in preserves 
%% -- il punto e' che, comunque cambi, QL resta!
We also note that introducing locality constraints based
on neighborhoods formally encompasses different specific notions that
are encountered in the physical and QIP literature
\cite{nielsen-chuang,Kraus-entangled,Verstraete2009}, where locality
is typically associated with operators of fixed maximum weight $t$
(also called ``$t$-body'' or ``$t$-local'' interactions) and/or
``distance'' on a graph.  For instance, when neighborhoods coincide
with individual subsystems, strictly local (or ``single-site'')
dynamics is enforced, with the corresponding propagator ${\cal T}_t$
being fully factorized relative to the multipartite structure, $ {\cal
T}_t \equiv \otimes_{a=1}^n {\cal T}_{a,t}$ \cite{ticozzi-QDS}.
Likewise, for $t=2$, allowing for arbitrary two-body (possibly
long-range) interactions correspond to identifying neighborhoods with
all possible subsystem pairs, whereas interactions between
nearest-neighbors sites may be accounted for by restricting the
neighborhoods accordingly.

\subsection{Stability definitions and standard form for stabilization}
\label{standardf}

As mentioned, our focus in this paper is on stabilizing a desired 
\emph{pure} state of $S$.  The relevant notions of stability are 
given in the following:

\begin{defin}\label{def:stability}
Let $\rho_d=\ket{\Psi}\bra{\Psi}\in{\mathfrak D}(\Hi)$ be a pure state
of $S$ evolving under Eq. \eqref{eq:lindblad}.

(i) $\rho_d$ is {\em Globally Asymptotically Stable (GAS)} if for
every initial condition $\rho_0 \in {\mathfrak D}(\Hi)$ we have
\begin{equation}
\lim_{t\rightarrow +\infty} e^{{\cal L}t}[\rho_0]= \rho_d. 
\label{eq:gas}
\end{equation}

(ii) If $\Hi'\subsetneq \Hi$ is a proper subspace of $\Hi$, $\rho_d$
is {\em Conditionally Asymptotically Stable relative to $\Hi'$} (or
simply $\Hi'$-{\em AS}) if Eq. (\ref{eq:gas}) holds for
$\rho_0 \in {\mathfrak D}(\Hi')$.
\end{defin}

A necessary condition for a state to be GAS is that it is {\em
invariant} \footnote{This follows directly from the fact that
Eq. \eqref{eq:lindblad} defines a \emph{linear autonomous} dynamical system,
as long as the generator ${\cal L}$ is {\em time-independent} as
assumed.  For more general dynamical systems, different notions of
stability may be relevant, see {\em e.g.} Ref. \cite{LaSalle}.},
namely, that it is in the kernel of the Liouvillian, ${\cal
L}(\rho_d)=0$.  The following proposition, that follows from the
general results of \cite{ticozzi-QDS}, provides a particularly simple
way to check for invariance if $\rho_d$ is pure:

\begin{Prop}
\label{prop1} 
Let the dynamics be driven by ${\cal L} (H, \{L_k\})$ as in
Eq. (\ref{eq:lindblad}). Then a state
$\rho_d=\ket{\Psi}\bra{\Psi}\in{\mathfrak D}(\Hi)$ is invariant if and
only if \beqa
\label{cond1}
&L_k\ket{\Psi} = \ell_k
\ket{\Psi},\quad \ell_k\in\C , \;\forall k,\\ 
&\tilde H\ket{\Psi} = h
\ket{\Psi}, \quad \tilde H=H - \frac{i}{2}\sum_k
(\ell_k^*L_k-\ell_kL_k^\dag), \; h\in\R.
\label{cond2} 
\eeqa 
\end{Prop} 

\vspace*{12pt}
\noindent
{\bf Proof:} Let $\Hi_d \equiv \span\{{\ket{\Psi}}\}$ and, with
respect to the natural  representation induced by the partition 
$\Hi =\Hi_d\oplus\Hi_d^\perp$, let operators be given in the block 
form 
\[ X=\left[ \begin{array}{c|c}
  X_d & X_P     \\\hline
  X_Q & X_R   
\end{array}
\right], \] where the top-left block is one-dimensional.  Given
Corollary 1 in \cite{ticozzi-QDS}, it follows that $\rho_d$ is
invariant if and only if $L_k |\Psi\rangle = l_{k,d} |\Psi\rangle
\equiv l_k |\Psi\rangle $ for all $k$, and the entries in the
$P$-block obey $i H_P - (1/2)\sum_k \ell_k^\ast L_{P,k}=0$. Clearly,
if $\ell_k \equiv 0$ in Eq. (\ref{cond1}), then it follows that
$H_P=0=H_Q$ by Hermiticity.  Therefore, $H \equiv \tilde{H}$ is
block-diagonal, with $H \ket{\Psi} = h_d \ket{\Psi}\equiv h_d
\ket{\Psi}$, as stated in (\ref{cond2}).  If $\ell_k \ne 0$ for some
$k$, we can use Eq. (\ref{eq:equiv}) to redefine $\tilde{L}_k\equiv
{L}_k -\ell_k I$ and, accordingly, $\tilde{H}$ as in (\ref{cond2}).
Since now $\tilde{\ell}_k \equiv 0$, $\tilde{H}$ is, again,
block-diagonal, with $\ket{\Psi}$ being an eigenstate with eigenvalue
$\tilde{h}_d\equiv h$, as stated.  \qed

Thanks to the above result, it is always possible 
to express a Markovian generator that asymptotically stabilizes a pure 
state in a {\em standard form},
in which the desired target state $\ket{\Psi}$ is annihilated by all
the noise operators (thus being a ``dark state'' in quantum-optics
language \cite{Kraus-entangled}) and is an eigenstate of the
corresponding Hamiltonian.  That is:

\begin{cor}
\label{standardform} 
If a generator ${\cal L}(H, \{ L_k\})$ makes
$\rho_d=\ket{\Psi}\bra{\Psi}\in{\mathfrak D}(\Hi)$ GAS, then the same
generator can be represented in a standard form ${\cal L}(\tilde{H},
\{ \tilde{L}_k \})$, in such a way that
$\tilde{H}\ket{\Psi}=h\ket{\Psi}$ and $\tilde{L}_k\ket{\Psi}=0$, for
all $k$.
\end{cor}

\subsection{Quasi-local stabilization settings}
\label{sec:settings}

In general, a QL stabilization problem will entail the specification
of the desired target state and the relevant QL constraints, along
with a description of the target dynamical model and control
parameters.  While, as mentioned, we shall focus here on pure target
states, different stabilization settings may be envisioned depending
on the nature of the available control resources and initialization
capabilities, and on the existence of non-trivial Markovian dynamics
in the absence of control, hereby referred to as \emph{drift} in the
standard control-theory terminology and characterized by a Liouvillian
generator ${\mathcal L}_0$.  Specifically, let us consider a class of
dynamical models as in Eq. (\ref{eq:lindblad}), with the corresponding
generator ${\mathcal L}\equiv {\mathcal L} ({H}, \{ {L}_k\})$ {\em
taken to be in standard form henceforth}, and
\begin{equation}
\dot{\rho}(t)={\mathcal L}(\rho(t)) \equiv \left [{\mathcal L_0} + 
{\mathcal L_c}\right] (\rho(t)).
\label{eq:drift}
\end{equation}
Here, ${\mathcal L}_0 ({H}_0, \{ L^0_k\})$ and ${\mathcal L}_c ({H}_c,
\{ D_k\})$ are associated to the drift and control dynamics,
respectively, and $H=H_0+{H}_c$, $\{ L_k \} = \{ L^0_k, D_k\}$.
Hamiltonian control is introduced by the time-independent $H_c$,
%% LV: Avevo usato 'coherent' control for i fisici, ovviamente... 
whereas each of the Lindblad operators $D_k$ represent a different
incoherent control knob on $S$. We can then introduce the following
definitions.

\begin{defin} 
A pure state $\rho_d=\ket{\Psi}\bra{\Psi}\in{\mathfrak D}(\Hi)$ is:

(i) {\em Quasi-Locally Stabilizable (QLS)} if there exist
a QL Hamiltonian $H_c$ and QL noise operators $\{
D_k\}_{k=1,\ldots,K}$ in standard form, such that $\rho_d$ is GAS for
the controlled Lindblad evolution in Eq. (\ref{eq:drift}).

(ii) {\em Dissipatively Quasi-Locally Stabilizable (DQLS)}
if it is QLS with ${H}_c\equiv 0$ and QL noise operators $\{D_k\}_{k=1,\ldots,K}$ in standard form.
%FT-simplified:"if there exist QL noise operators $\{D_k\}_{k=1,\ldots,K}$ in standard
%form, such that $\rho_d$ is GAS for the dissipatively controlled
%Lindblad evolution in Eq. (\ref{eq:drift}) (${H}_c\equiv 0$)."
%% LV: Sure!

(iii) {\em Conditionally DQLS with respect to a subspace}
$\Hi'\subsetneq \Hi$ (or simply $\Hi'$-{\em DQLS}), if there exist QL
noise operators $\{D_k\}_{k=1,\ldots,K}$ in standard form, such that
$\rho_d$ is 
%{\ft $\Hi'$-AS} 
conditionally AS relative to $\Hi'$ 
%% LV: preferirei essere + esplicita
for the dissipatively controlled Lindblad
evolution in Eq. (\ref{eq:drift}).
\end{defin}

Two features follow from the global stability properties given in
Definition \ref{def:stability}: while {\em exact} QL %FT: questa e` un'altra cosa:"steering" 
%LV: Ok scusa!...
preparation cannot be achieved in finite time, convergence 
%FT: no "generally", sempre, asintoticamente... e` lineare.
%%  LV: E gli autovalori zero?.... 
%
happens exponentially fast in time \cite{ticozzi-NV}.
Furthermore, if the target state is stabilizable, according to the
appropriate definition, the stabilizing dynamics is {\em robust with 
respect to the initial state} \cite{ticozzi-QDS} thus, equivalently, the 
control parameters are {\em all-to-one} \cite{Pechen}. 
%%FT: Evvabbe`, non l'hanno mica inventata loro l'analisi di stabilita`!
%% rimosso e sostituito: {control parameters are, by
%% construction, {\em all-to-one} \cite{Pechen}. } 
% LV: Si' pero' anche tu hai un po' la tendenza a citare preferibilmente i tuoi lavori ed evitare 
%       connessioni con quelli di altri..........
%
%FT: Qui non ho capito cosa vuoi dire -- optimality rispetto a che? Speed? Non mi % arrischierei di fare un claim cosi`...
%In principle, this
%allows for such controls to be simultaneously optimal for arbitrary
%initial states in $\Hi$ (or $\Hi'$ in the conditional case).
%% LV: No, se leggi il lavoro di Pechen, significa semplicemente che se trovi un controllo 
%%       che realizza la preparazione in modo ottimale (=minimizzando un certo funzionale), 
%%       te lo fa indipendentemente dallo stato iniziale

Clearly, the QLS setting includes both purely dissipative and
conditional stabilization as special instances where either (or both)
the admissible control actions and initial states are restricted,
respectively.  Introducing a separate DQLS notion is motivated by the
fact that it allows for a simpler mathematical treatment (see
Sec. \ref{prior}) while being adequate for important classes of
entangled pure states and, from a practical standpoint, potentially
simpler to implement.  For states which are not DQLS, QLS may or may
not be achievable. Even in the latter case, conditional stabilization
may offer a practical method of choice in situations where a two-step
``switched'' dynamics, that first initializes $S$ to the intended
subspace, and from there enacts dissipative stabilization, is
preferable over identifying and implementing a combined
Hamiltonian-dissipative control action -- for instance thanks to the
presence of a conserved quantity \cite{schindler}.  While a number of
illustrative examples will be discussed in Sec. \ref{sec:example},
ensuring that the required number $K$ of noise operators scales
favorably with the complexity of the neighborhood structure is crucial
in practice.  The situation is straightforward for the DQLS setting,
since it follows directly from Theorem \ref{mainthm} below that a {\em
single} noise operator per neighborhood always suffices ($K=M$).
In general, the randomized algorithms that we will provide to achieve
stabilization in case (i) and (ii) will also output, when successful,
a noise operator per neighborhood, as we shall see.

We begin by presenting two preliminary results that directly
generalize their counterparts in Ref. \cite{ticozzi-ql} for the DQLS
case.  First, let $U=\bigotimes_{a=1}^n U_a$ be an arbitrary local
unitary (LU) transformation.  Then it is straightforward to see that
the above stabilization notions are invariant under arbitrary LU
transformations of the target state, as desirable given that
entanglement properties are themselves preserved under LUs.  The
following Proposition may be established through the same steps used
in Lemma 2.2 and Proposition 2.3 in \cite{ticozzi-ql}:

\begin{Prop}
\label{lu} 
If $\rho_d $ is QLS and $U$ is any LU, then $\rho'_d= U \rho_d
U^\dagger$ is also QLS. If $\rho_d $ is $\Hi'$-DQLS and $U$ is any LU
that leaves $\Hi'$ invariant, then $\rho'_d= U \rho_d U^\dagger$ is
also $\Hi'$-DQLS.
\end{Prop}

A second basic yet useful result regards the structure of the
operators $\{ D_k\}$: the Lemma that follows shows that the support of
a QLS state must still be, as in the DQLS case, contained in the
kernel of the noise operators written in standard form:

\begin{lemma}
\label{supp} Assume that the generator associated to QL
$\{ H, D_k\}$ leaves $\rho_d=\ket{\Psi}\bra{\Psi}$
invariant. Then, for each $k,$ we have $\supp(\rho_{{\cal
N}_k})\subseteq \ker( D_{{\cal N}_k}).$
\end{lemma}

\vspace*{12pt}
\noindent
{\bf Proof:} If the state is invariant, by Proposition \ref{prop1},
$\ket{\Psi}$ must be in the kernel of each $D_k$. Thus, with respect
to the decomposition $\Hi =\Hi_d \oplus\Hi_d^\perp,$ with
$\Hi_d=\textrm{span}\{\ket{\Psi}\}$ as before, every $D_k$ must be of
block form \cite{ticozzi-markovian}:
\[D_k=\left[\begin{array}{cc} 0 & D_{P,k} \\ 0 & D_{R,k}
\end{array}\right],\]
which immediately implies $D_k\rho_d D_k^\dag =0.$ It then follows
that $\tr_{\bar{\cal N}_k}$ $(D_k\rho_d D_k^\dag)=0$, therefore
$\tr_{\bar{\cal N}_k}$ $( D_{{\cal N}_k}\otimes I_{\bar{\cal
N}_k}\rho_d D_{{\cal N}_k}^\dag\otimes I_{\bar{\cal N}_k})=0$.  Thus,
it also follows that $D_{{\cal N}_k}\rho_{{\cal N}_k} D_{{\cal
N}_k}^\dag=0.  $ If we consider the spectral decomposition
$\rho_{{\cal N}_k} \equiv \sum_jq_j\ket{\phi_j}\bra{\phi_j}$, with
$q_j>0,$ the latter implies that, for each $j,$ $\tilde D_{{\cal
N}_k}\ket{\phi_j}\bra{\phi_j}\tilde D_{{\cal N}_k}^\dag=0$.  Thus, it
must be $\supp(\rho_{{\cal N}_k})\subseteq \ker(\tilde D_{{\cal
N}_k})$, as stated. \qed

For added clarity and notational simplicity, we shall from now on
assume that \emph{the dynamics of the system is drift-less}, that is,
${\cal L}_0\equiv 0$ in Eq. (\ref{eq:drift}).  After briefly recalling
the main results on DQLS, we
will present our new results on Hamiltonian-assisted and conditional
QL stabilization under the simplyfing drift-less assumption, in
Sec. \ref{sec:qls} and Sec. \ref{sec:conditional}, respectively.  We
will then allow for drift dynamics and explicitly address its role
in Sec. \ref{sec:drift}.

\subsection{Prior results for quasi-local dissipative stabilization}
\label{prior}

A characterization of DQLS states, leading to a simple
\emph{linear-algebraic algorithm} to test whether a given pure state
is DQLS, may be obtained based on the properties of the reduced states
on the neighborhoods~\footnote{Note that in this simple setting, if
the target state admits a partial factorization ({\em e.g.}, a product
of two entangled states), there is no loss of generality in
stabilizing such factors individually, possibly at the expenses of
redefining the neighborhood structure to make it compatible with the
given factorization, as noted in \cite{ticozzi-ql}.}, as defined in
Eq. (\ref{redstate}).  Let 
\beq 
{\mathcal H}^\circ_{k}\equiv \text{supp}(\rho_{{\cal N}_k}
\otimes I_{\bar{\cal N}_k}), \hspace*{1cm} 
{\mathcal H}_0 = \bigcap_k {\mathcal H}^\circ_{k}.
\label{Hzero}
\eeq
\noindent 
We thus have the following:

\begin{thm}
\label{mainthm} 
{\bf \cite{ticozzi-ql} } A pure state $\rho_d=|\Psi\rangle\langle
\Psi|$ is DQLS if and only if 
\beq
\label{GAScond} 
\text{supp}(\rho_d) = {\mathcal H}_0.\eeq
\end{thm}

The proof of this Theorem includes an explicit construction of a
choice of stabilizing noise operators, with their general
block-structure highlighted in Lemma \ref{supp}. It also points to
natural connections with the formalism of {\em parent Hamiltonians}
and the concept of {\em frustration} from many-body physics.  Let
$H=\sum_k H_k$, with $H_k=H_{{\cal N}_k}\otimes I_{\bar{\cal N}_k}$
and, as before, the index $k$ refers to the $k$th neighborhood.  A QL
Hamiltonian $H$ is said to be a {\em parent Hamiltonian} for $|\Psi
\rangle$, if $|\Psi \rangle$ is an {\em exact} ground state for $H$
\cite{Fannes}.  In addition, $|\Psi \rangle$ is said to be {\em
frustration-free} if it is the exact ground state of each Hamiltonian
$H_k$ separately, that is, $\langle \Psi |H_k | \Psi
\rangle=\textrm{min}\; \lambda(H_k)$, $\forall k,$ where
$\lambda(\cdot)$ denotes the spectrum of a matrix.  Remarkably, it is
known that for a large class of product entangled-pair states (PEPSs)
or, in one spatial dimension, matrix product states (MPSs), a frustration-free 
parent Hamiltonian may be constructed\footnote{This class includes all
so-called (blocked) {\em injective} PEPSs or MPSs. A non-injective MPSs 
is the degenerate ground state of a frustration-free parent Hamiltonian if 
all the matrices in the MPS description can be block-diagonalized to a form 
where each block corresponds to an injective MPS. However, there exists
non-injective MPSs that are still the unique ground state of their parent 
Hamiltonian.}{ }, which has the desired state as its {\em unique} ground state
\cite{perez-MPS,perez-PEPS,Norbert}. The QL properties of this Hamiltonian 
are determined by the (minimum) ``bond dimension'' of the corresponding
PEPS or MPS representation.
Suppose that a pure state is the unique ground state of a
frustration-free parent Hamiltonian.  Then the QL structure of $H$ may
be naturally used to derive a stabilizing semigroup
\cite{Kraus-entangled}. It is easy to show that this condition is also
necessary, leading to the following:

\vspace*{2mm}

\begin{cor}
\label{parent}
{\bf \cite{ticozzi-ql}} A state $\rho_d=|\Psi\rangle\langle\Psi|$ is
DQLS if and only if it is the unique ground state of a
frustration-free parent Hamiltonian.
\end{cor}

Despite the formal points of contact, it is crucial to remark that in
MPS-based stabilization approaches \cite{Kraus-entangled} the relevant
QL notion is intrinsically {\em state-dependent}, whereas it is taken
to be a {\em fixed} problem input in our control-motivated approach.
From a practical standpoint, the DQLS class includes important
representative sets of entangled states -- most notably, {\em all
stabilizer and graph states} \cite{Kraus-entangled,ticozzi-ql},
relative to the natural choice of neighborhoods associated with
connected nodes on the graph.
Still, paradigmatic examples of genuinely multipartite entangled
states such as GHZ and W states can be easily seen to {\em fail} the
DQLS test in Eq. (\ref{GAScond}), except in ``fully connected'' (for
example, ``star'') coupling topologies where the QL constraint becomes
effectively trivial.  While we defer to \cite{ticozzi-ql} and
Sec. \ref{sec:example} for further discussion, we shall proceed to
separately formalize and analyze the general QLS and the conditional
DQLS scenarios next. 

%---------------------------------------------------------------------------

\section{Quasi-local stabilization with constant dissipative and 
Hamiltonian control}
\label{sec:qls}

\subsection{Necessary conditions}

Assume that the target state $\rho_d=|\Psi\rangle\langle\Psi|$ is not
DQLS, and let $\Hi_d=\span\{|\Psi\rangle\}$ as before, corresponding
to the support of the desired state.  In terms of $\Hi_0,$ defined as in
Eq. (\ref{Hzero}), this translates to:
\[\textrm{dim}\left(\Hi_0\right)= 
\textrm{dim}\Big(\bigcap_k {\mathcal H}_k^\circ \Big)\equiv d_0\geq
2. \]
\noindent 
Our first result is the following necessary condition for QLS (recall 
that $H\equiv H_c$ in the drift-less scenario we consider for now): 

\begin{Prop}
\label{effectiveH}
$\rho_d$ is QLS but not DQLS only if $\Hi_d$ is an invariant subspace
for the Hamiltonian $H_c$ and {\em no} other invariant subspace is
contained in (or equal to) $\Hi_0.$ In particular, one can choose $H_c$
so that $H_c \ket{\Psi} = 0.$
\end{Prop}
%% LV: Ho cambiato con H_c per consistenza con (10) below

\vspace*{12pt}
\noindent
{\bf Proof:} If there were another invariant subspace for the
Hamiltonian with support in $\Hi_0,$ the latter would be, by
definition of $\Hi_0,$ also in the kernel of each $D_k,$ and hence it
would be invariant. Invariant subspaces always contain at least an
invariant state. It then follows that $\rho_d$ could not be GAS.  If
$H_c \ket{\Psi}=\lambda\ket{\Psi},\,\lambda\neq 0$, we can always
choose $ H'_c=H_c-\lambda I$ instead, which is also QL if $H_c$ was.
\qed

When such a stabilizing Hamiltonian $H_c$ exists, one must look for
noise operators $\{D_k\}$ such that $\Hi_d$ is the {\em only}
invariant subspace for the whole generator ${\cal L}_c (H_c, \{D_k\})$,
and thus makes $\rho_d$ GAS. Lemma \ref{supp} suggests that the most
effective choice of noise operators can stabilize $\Hi_0,$ but do no
better than that. In order to specify what the action of a stabilizing
Hamiltonian would be, it is convenient to pick an orthonormal basis
for $\Hi_0,$ which includes the target state:
\[\Hi_0=\textrm{span}\{\ket{\Psi},\ket{\Phi_1},\dots,\ket{\Phi_r}\},\quad
r=d_0-1.\]
\noindent 
One would hope that $H_c \ket{\Phi_j}\notin\Hi_0$ for each $j.$
However, fulfilling these conditions is clearly not necessary, and in
fact it need not be possible given the QL constraint.  However, in the simplest
case of $d_0=2,$ the above idea leads to a specialized formulation
of Proposition \ref{effectiveH}:

\begin{cor}
\label{effectiveH2}
$\rho_d$ is QLS but not DQLS, with
$\Hi_0=\text{span}\{\ket{\Psi},\ket{\Phi_1}\},$ only if there exists
a QL Hamiltonian $H_c$ such that \beq H_c \ket{\Psi} = 0,
\hspace*{1cm} H_c \ket{\Phi_1} \notin \Hi_0.  \eeq
\end{cor}
%% LV: Fixed inconsistency between the two eqs!

Recall that in the DQLS context (Corollary \ref{parent}), a QL parent
Hamiltonian was naturally associated to the action of the noise
operators, in such a way that $\Hi_0$ was the common ground eigenspace
of all the QL components, that is, with no frustration involved.  In
contrast, the necessary conditions provided above clearly show that an
effective control Hamiltonian $H_c$ {\em cannot be frustration-free}
with respect to $\Hi_0$, since it must destabilize some part of
$\Hi_0$ in order to attain GAS of the target state.

In Sec. \ref{sec:example} we will employ Corollary \ref{effectiveH2}
to construct a stabilizing Hamiltonian for both GHZ and W states
which, as noted earlier, are {\em never} DQLS under non-trivial QL
constraints.

\subsection{Randomized construction of stabilizing Hamiltonian and dissipators}
\label{sec:random}

Even if we succeed in finding a Hamiltonian $H_c$ that satisfies
Proposition \ref{effectiveH} (or Corollary \ref{effectiveH2}), a
procedure for determining the existence and, possibly, the actual form
of the stabilizing $D_k$ is required in order to establish QLS.  As we
will illustrate with an example in Sec. \ref{GHZ}, {\em not all
choices of $\{D_k\}$ satisfying Lemma \ref{supp} are effective}. In
fact, care is needed in ensuring that the interplay between
Hamiltonian and dissipative control introduces enough ``mixing'' and
does not allow for other invariant sets to exist.

With that in mind, we shall invoke a {\em randomized} design approach
that expands our earlier use of randomization in \cite{ticozzi-NV} to
{\em both} the Hamiltonian and the dissipative component, and prove
that a {\em generic} choice of noise operators which satisfy Lemma
\ref{supp} and stabilize $\Hi_0$ will suffice to achieve
QLS. Specifically, assume that we represent the desired stabilizing
operators $H_c = \sum_k H_k$, and $\{D_k\}$ in parametric form: 
\beq
H_k= \sum_j \alpha_{jk}\sigma_{jk},\quad
D_k=\sum_j\beta_{jk}\sigma_{jk},
\label{param}
\eeq 
\noindent 
where $\alpha_{jk},\beta_{jk} \in {\mathbb R}$ are chosen at random
with uniform distribution in an interval ${\cal I}\equiv [-\gamma,\gamma]$
%FT:I e` l'identita`.. 
% LV: OOps!  
of the real axis, and $\{\sigma_{jk}\}$ is a basis$\,$\footnote{Since we
consider real linear combinations, the dimension is doubled with
respect to the more usual complex field, see Appendix B.}$\,${ } for the
space of QL operators on the $k$th neighborhood ${\cal N}_k$. Our main
result is then contained in the following:

\begin{thm}
\label{randomized}
If there exists a choice of $\alpha_{jk},\beta_{jk}\in {\cal I}$ that makes
$\rho_d$ QLS, then {almost any} choice of $\alpha_{jk},\beta_{jk}
\in I$, that makes $\rho_d$ invariant, makes it QLS as well.
\end{thm}

While the proof of the above Theorem is rather technical (see Appendix
\ref{sec:randomized}), the meaning is clear: imposing invariance
requires ``fine-tuning'' of the parameters, however once invariance is
ensured, \emph{if} GAS is possible, then it comes almost always for
free.  An algorithm for constructing a stabilizing QL Hamiltonian and
achieve QLS may then be provided as follows.

\begin{description}

\item{Step 1:} {\em Imposing quasi-locality of $\{H_c\}$}. Pick a
product operator basis for the $d^2$-dimensional space of linear
operators ${\mathfrak B}(\Hi),$ say,
$\{\sigma_{i_1}\otimes\ldots\otimes\sigma_{i_n}\,|\,$$
{i_a=1,\ldots,d_a^2}\}$.  Let $H_c=\sum_k H_k,$ and associate each
$H_k$ and $D_k$ to a vector $\vec h_k$ and $\vec d_k$, respectively.
Let $\hat D=[\,\vec d_1|\cdots|\vec d_n\,]$ be the matrix of the
coefficient of the noise operators relative to the above product
basis, and $\hat B_{k}$ the orthogonal projection onto the subspace
generated by the basis elements that are QL with respect to ${\cal
N}_k$, with $\hat B_{k}^\perp=I-\hat B_k$.  We must then require, for
each $k$:
\[\hat B_{k}^\perp\vec h_k =0, \quad \hat B_{k}^\perp\vec d_k =0.\]

\item{Step 2:} {\em Ensuring invariance of $\Hi_0$}. Impose the linear
constraints on the QL noise operators and on the Hamiltonian, namely
$D_k\ket{\psi}=0,$ $H_c\ket{\psi}=0$ for all $k$. If $\hat P_0$ is the
matrix representation of the linear (super)-operator
$P_0(X)=X\ket{\psi}\bra{\psi}$ with respect to the the chosen basis,
this is equivalent to require that for each $k$, 

\[\hat P_0\vec h=0,\quad \hat P_0\vec d_k=0.\]

These two steps translate, for each $k$, in the following homogeneous
systems of linear equations: \beq
\label{linsys2} 
\left[ \begin{array}{c}\hat B_{k}^\perp\\\hat
P_0\end{array}\right]\vec d_k=0,\quad \left[ \begin{array}{c}\hat
B_{k}^\perp\\\hat P_0\end{array}\right]\vec h_k=0 . 
\eeq

\item{Step 3:} {\em Enforcing convergence by randomization}. If the
constraints above allow for non-trivial solutions (and hence a
subspace of solutions since the system is homogenous), choose the free
variables uniformly at random within a finite interval
$[-\gamma,\gamma] \subset {\mathbb R}$.

\end{description}

\noindent 
We thus have the following immediate corollary of Theorem
\ref{randomized}:

\begin{cor} If $\rho_d$ is QLS, the QL generator ${\cal L}_c (H_c, \{D_k\})$ 
constructed in Steps 1--3 makes it GAS with probability one. If
$\rho_d$ is not QLS, then it is not stabilized by the constructed
Hamiltonian $H_c$ and noise operators $\{D_k\}$.
\end{cor}

Since checking whether a state is the unique equilibrium of some
Lindblad dynamics is straightforward (\emph{e.g.}, by checking that
the superoperator form of the corresponding generator has a unique
unit eigenvalue), the above algorithm can be used as an explicit test
for QLS.
%%% MEMO: How many Lindblads?

%----------------------------------------------------------------------------

\section{Conditional quasi-local dissipative stabilization}
\label{sec:conditional}

\subsection{Characterization of conditional stability}

When a state is not DQLS and Hamiltonian control is not viable,
we can return to a purely dissipative control setting and analyze our
second proposed stabilization strategy, namely to restrict the initial state 
to a given initial subspace of $\Hi$. With
$\Hi_d=\span\{|\Psi\rangle\}$ and $\Hi_0$ given in Eq. (\ref{Hzero})
as before, let us additionally define the following subspaces:
\[ \Hi_w=\Hi_0\ominus\Hi_d, \hspace*{1cm} \Hi_r=\Hi\ominus\Hi_0. \]
\noindent 
By construction, $\Hi_w$ corresponds to the portion of the subspace
stabilizable by purely dissipative means that is orthogonal to the
target.  Assume that the dynamics is given by Eq. (\ref{eq:drift}),
with ${\cal L}\equiv 0$, associated to QL noise operators $\{D_k\}$
that make ${\mathfrak D}(\Hi_0)$ GAS.  We aim to characterize which
choice(s) of $\Hi'$ can make $\rho_d$ conditionally DQLS (with the
obvious requirement that $\Hi'\supsetneq \Hi_d$). A first necessary
condition is provided by the following Lemma:

\begin{lemma}\label{Hperp} 
$\rho_d$ can be $\Hi'$-DQLS only if  
\beq
\label{h11}
\Hi'\subseteq\Hi\ominus\Hi_w = \Hi_d \oplus [\Hi \ominus \Hi_0].\eeq
\end{lemma}

\vspace*{12pt}
\noindent
{\bf Proof:} Since ${\mathfrak D}(\Hi_w)\subsetneq{\mathfrak
D}(\Hi_0),$ and by hypothesis ${\mathfrak D}(\Hi_0)$ is invariant
under the Lindblad dynamics ${\cal L}_D$ induced by the $\{D_k\}$,
then ${\mathfrak D}(\Hi_w)$ is invariant as well. If
$\Hi'\subseteq\Hi\ominus\Hi_w$ did not hold, then there would be a
$\rho \in {\mathfrak D}(\Hi')$ such that $\tr$ $(\rho\Pi_{\Hi_w})>0$,
with $\Pi_{\Hi_w}$ denoting the projector onto $\Hi_w$.  Define
$p_w=\tr$ $(\Pi_{\Hi_w}\rho)>0$,
$\rho_w=p_w^{-1}\Pi_{\Hi_w}\rho\Pi_{\Hi_w}\in{\mathfrak D}(\Hi_w),$
and write $\rho=p_w\rho_w + \Delta\rho,$ for some Hermitian but not
necessarily positive operator $\Delta\rho$. Since the dynamics is
linear and $\rho_w$ invariant, $\tr$$(\rho\Pi_{\Hi_w})$ cannot
decrease along the dynamical flow \cite{ticozzi-QDS}, therefore we may
write
\[\lim_{t\rightarrow\infty}\text{Tr}(e^{{\cal L}_Dt}[\rho]\Pi_{\Hi_w})=
\text{Tr} (e^{{\cal L}_Dt}[\Delta\rho])+ p_w\text{Tr}(\rho_w)\geq p_w
>0.\] Hence $\rho\in{\mathfrak D}(\Hi')$ would not converge to
$\rho_d.$ \qed

The following ``enlargement lemma'' indicates that we can 
construct larger stabilizing subspaces out of smaller ones, both weakening
our constraints and gaining a key property: {\em invariance}.  Formally:

\begin{lemma}
\label{Hinv} 
If $\rho_d$ is $\Hi''$-DQLS, it is also $\Hi'$-DQLS for some $\Hi'$
such that $\Hi''\subset\Hi'$ and ${\mathfrak D}(\Hi')$ is an invariant
set.
\end{lemma}

\vspace*{12pt}
\noindent
{\bf Proof:} In order to make ${\mathfrak D}(\Hi')$ invariant, it
suffices to combine the supports of all trajectories originated in
${\mathfrak D}(\Hi'')$. Given two subspaces $\Hi_{1,2}$, let us denote
by $\Hi_1\vee \Hi_2,$ the smallest subspace that contains both of
them. Define
\[ \Hi_{\rho_0}=\bigvee_{t\geq0}\supp(e^{{\cal L}_Dt}[\rho_0]), 
\hspace*{1cm} \Hi'=\bigvee_{\rho_0\in{\mathfrak
D}(\Hi'')}\Hi_{\rho_0}.\] 
\noindent 
By assumption, $\Hi_d\subsetneq\Hi'$, $\Hi''\subsetneq\Hi'$ and
${\mathfrak D}(\Hi')$ is invariant. Lastly, all trajectories
originated in ${\mathfrak D}(\Hi')$ can be obtained from trajectories
originated in ${\mathfrak D}(\Hi''),$ which by hypothesis all converge
to $\rho_d$, establishing the desired result. \qed

Motivated by the above, we now restrict to subspaces $\Hi'$ such that
${\mathfrak D}(\Hi')$ is invariant.  With a slight abuse in
terminology, we shall simply refer to subspaces obeying such a
property as invariant.  We are then ready to state our main result:

\begin{thm}
\label{conditionalstab} 
Assume that the QL dissipative dynamics generated by the $\{D_k\}$
makes ${\mathfrak D}(\Hi_0)$ GAS.  If $\Hi'$ satisfies the necessary
condition given in Eq. \eqref{h11} and is invariant for all $\{D_k\},$
then $\rho_d$ is $\Hi'$-DQLS.
\end{thm}

\vspace*{12pt}
\noindent
{\bf Proof:} Since $\Hi'$ is invariant for all $\{D_k\}$, it follows
that ${\mathfrak D}(\Hi')$ is a positive invariant set. By LaSalle
theorem \cite{LaSalle}, this means that all the trajectories starting
in ${\mathfrak D}(\Hi')$ converge to its largest invariant set.  On
the other hand, by hypothesis, ${\mathfrak D}(\Hi_0)$ is GAS, so the
largest invariant set must be contained in ${\mathfrak
D}(\Hi_0)\cap{\mathfrak D}(\Hi').$ Since by \eqref{h11} we know that
$\Hi_w\perp\Hi',$ it follows that
\[{\mathfrak D}(\Hi_0)\cap{\mathfrak D}(\Hi')={\mathfrak D}(\Hi_d)=\rho_d,\]
as desired. \qed

A further advantage of considering an invariant subspace $\Hi'$ is
highlighted in the following corollary: if initialization in $\Hi'$ is
faulty, the error on the asymptotic result remains upper-bounded by
the preparation error:

\begin{cor} 
Let $\rho_d$ be $\Hi'$-DQLS, with $\Hi'$ invariant under the
dissipative dynamics. If $\rho$ is such that
\[1-\tr(\Pi_{\Hi'}\rho) \equiv \varepsilon>0,\] then we have:
\[\lim_{t\rightarrow\infty}\tr(e^{{\cal L}_Dt}[\rho]\Pi_{\Hi'})<
\varepsilon .\]
\end{cor}

\vspace*{12pt}
\noindent
{\bf Proof:} Since $\Hi'$ is invariant, $\tr(\Pi_{\Hi'}\rho)$ is non
decreasing along the trajectories $e^{{\cal L}t}\rho,$ for $t\geq 0$, 
see \cite{ticozzi-QDS}.  
\qed

While it seems hard to devise a fully general strategy for finding
good choices of $\Hi'$ and associated QL dissipators so that $\rho_d$
is $\Hi'$-DQLS, constructive results may be obtained if the problem is
further constrained.  In particular, note that from a stabilization
point of view, two subspace decompositions play a specially important
role: 

(i) one associated with the \emph{initial} ($t=0$) state-space
structure, $\Hi=\Hi' \oplus \Hi'^\perp$, with $\Hi'$ containing the
states to be attracted toward $\rho_d$; 

(ii) one associated with the {\em final} ($t \rightarrow \infty$)
state-space structure, $\Hi_0= \Hi_d \oplus \Hi_w$, with $\Hi_w$
containing the unwanted states toward which dissipative stabilization
occurs.

\noindent 
In a way, the two classes of conditional stabilization problems we
solve can be seen to arise by imposing some natural constraints on
$\Hi'^\perp$ and $\Hi_w$, in case (i) and (ii), respectively.  Either way, 
it is worth remarking that {\em any state in ${\mathfrak D}(\Hi')$ is 
asymptotically converging to $\rho_d$:} if control capabilities are
enough to prepare a subset or even a single state in this set, dissipative 
QL preparation can be achieved.  We begin by addressing case (ii), which 
is directly motivated by W states and is technically simpler.

%{\ft Finally, it is worth remarking that {\em any state in ${\mathfrak D}(\Hi')$ is asymptotically converging to $\rho_d$:} if my control capabilities are enough to prepare a subset or even a single state in this set, dissipative quasi-local preparation can be achieved.}
%%LV: Ok ma ho spostato

\subsection{Conditional stabilization under constraints on the 
final attractive set} 
\label{wtype}

In order to formulate and interpret the required property that $\Hi_w$
must obey to allow for $\Hi'$ and associated QL dissipators to be
systematically constructed, some additional definitions are needed.
Let
\[ \Hi^\circ_{{\cal N}_k}=\supp(\rho_{{\cal N}_k}),\hspace*{5mm}
\Hi^r_{{\cal N}_k}=\Hi_{{\cal N}_k}\ominus\Hi^\circ_{{\cal N}_k}, \]
%\hspace*{5mm} 
\[\Hi^w_{{\cal N}_k}=\supp\left( \text{Tr}_{{\bar{\cal
N}_k}}(\Pi_{\Hi_w})\right),\]
\noindent 
where as before $\Pi_{\Hi_w}$ is the orthogonal projector on $\Hi_w.$
By construction, $\Hi_w\subseteq\Hi_0$ and $\Hi^w_{{\cal
N}_k}\subseteq \Hi^\circ_{{\cal N}_k}$.  As we shall establish, the
key property that $\Hi_w$ must obey is the following {\em strict}
inclusion for each neighborhood:
\beq \Hi^w_{{\cal N}_k}\subsetneq \Hi^\circ_{{\cal N}_k}.
\label{nonover}\eeq

Some intuition on the above requirement may be built as follows. We
know that $\Hi_{{\cal N}_k}$ must be in the kernel of all the noise
operators $D_k$, and hence the dynamics in this subspace is
trivial. We thus would like to be able to distinguish $\Hi_d$ from
$\Hi_w$, and ``push'' the evolution towards the former but not the
latter.  If Eq. \eqref{nonover} is obeyed, we know that if a pure
state is in $\Hi_0$ but has {\em no QL support on any of the}
$\Hi^w_{{\cal N}_k},$ then it must be in $\Hi_d$ and hence in
$\rho_d$.  Formalizing this intution provides us with a way to
construct a suitable subspace $\Hi'$ and utilize Theorem
\ref{conditionalstab} to prove convergence.

Specifically, our candidate subspace for conditional invariance is
defined as follows:
\[ \Hi' \equiv \Hi\ominus\bigcap_k\tilde\Hi_k , \hspace*{1cm}
\tilde \Hi_k=\Hi^w_{{\cal N}_k}\otimes\Hi_{\bar{\cal N}_k} .\]
\noindent 
By Lemma \ref{supp} and the definition of $\Hi^w_{{\cal N}_k}$, $
\Hi_w\subseteq \bigcap_k \tilde \Hi_k.$ Hence, $\Hi'$ obeys the 
necessary condition established in Eq. \eqref{h11}:
\beq\label{st0}
\Hi'\subseteq \Hi \ominus \Hi_w.\eeq

We now need to construct noise operators $D_k=D_{{\cal N}_k}\otimes
I_{\bar{\cal N}_k}$, such that $\Hi_0$ and $\Hi'$ are invariant.
To this aim, define $\Hi^t_{{\cal N}_k} \equiv \Hi^\circ_{{\cal
N}_k}\ominus \Hi^w_{{\cal N}_k},$ which by Eq. \eqref{nonover} is not
empty, and consider $D_{{\cal N}_k}$ with the following block
structure, with respect to the decomposition of the QL space
$\Hi_{{\cal N}_k}=\Hi^w_{{\cal N}_k}\oplus\Hi^t_{{\cal
N}_k}\oplus\Hi^r_{{\cal N}_k}$:

\beq
\label{str}
D_{{\cal N}_k}=\left[\begin{array}{ccc} 0 & 0 & 0\\ 0 & 0 & D_{P,{\cal
N}_k} \\ 0 & 0 & D_{R,{\cal N}_k}
\end{array}\right].
\eeq The above structure can be exploited, {\em e.g.} via a choice of
$D_{P,{\cal N}_k},D_{R,{\cal N}_k}$ of ``ladder form''
\cite{ticozzi-QDS}, to render {\em each} of the $\Hi^\circ_{{\cal
N}_k} $ (hence $\Hi_0$) GAS. We hereby consider this or an equivalent
choice, so that $\Hi_0$ is made GAS with QL operators. In addition to
this, the block structure in Eq. \eqref{str} ensures the required
invariance property of $\Hi'$:

\begin{Prop}
A choice of $D_k$ satisfying Eq. \eqref{str} ensures that
$\Hi'=\Hi\ominus\bigcap_k\tilde\Hi_k$ is invariant.
\end{Prop}

\vspace*{12pt}
\noindent
{\bf Proof:} Given the matrix structure, we have:
\beqa
\label{st1}
&&D_k\Hi'\subseteq D_k(\Hi\ominus\tilde\Hi_k)\nonumber\\&&=(D_{{\cal
N}_k}\otimes I_{\bar{\cal N}_k})\left((\Hi^t_{{\cal
N}_k}\oplus\Hi_{{\cal N}_k}^r)\otimes\Hi_{\bar{\cal
N}_k}\right)\\&&\subseteq (\Hi^t_{{\cal N}_k}\oplus\Hi_{{\cal
N}_k}^r)\otimes\Hi_{\bar{\cal N}_k}.\nonumber\eeqa
\noindent 
Furthermore, 
\beq
\label{st2}
\bigcap_k\tilde\Hi_k\subset\tilde\Hi_k=\Hi^w_{{\cal
N}_k}\otimes\Hi_{\bar{\cal N}_k} \perp (\Hi^t_{{\cal
N}_k}\oplus\Hi_{{\cal N}_k}^r)\otimes\Hi_{\bar{\cal N}_k}.\eeq

Thus, by combining Eqs. \eqref{st1} and \eqref{st2}, we obtain
$D_k\Hi'\perp \bigcap_k\tilde\Hi_k,$ so by definition of $\Hi'$ we
have $D_k\Hi'\subseteq \Hi'.$ \qed

Given Eq. \eqref{st0} and the above proposition, we can then apply
Theorem \ref{conditionalstab} to establish that the constructed
dynamics makes $\rho_d$ conditional $\Hi'$-DQLS. We summarize the
results of this section in the following:

\begin{cor} 
\label{corwtype} 
Let $\rho_d$ and the given neighborhood structure $\{ {\cal N}_k\}$ be
such that Eq. \eqref{nonover} holds.  Then by choosing $\Hi'=
\Hi\ominus\bigcap_k ( \Hi^w_{{\cal N}_k}\otimes\Hi_{\bar{\cal N}_k}
)$, and a set of QL $\{D_k\}$ that satisfy Eq. \eqref{str},
%such that $\Hi_0$ is GAS. Then 
%% LV: Serve specificarlo nell'enunciato?
$\rho_d$ is $\Hi'$-DQLS.
\end{cor}

As remarked, a notable example of states satisfying the property in
Eq. \eqref{nonover} is the class of W states. Their stabilization
using this technique will be explicitly addressed in Sec.
\ref{sec:example}.

\subsection{Conditional stabilization under constraints on the 
initial attraction basin}
\label{initialbasin}

In this case, the additional assumption we impose on the noise
operators $\{D_k\}$ is that {\em both $\Hi'$ and $\Hi'^\perp$ are left
invariant}.
%We here provide the main steps of a strategy for testing $\Hi'$ and
%constructing an effective set of noise operators $\{D_k\}$ that
%simultaneously leave $\Hi'$ {\em and} $\Hi'_R:=\Hi\ominus\Hi'$
%invariant.
%% LV: \Hi'_R vs. \Hi'^\perp --- capisco le ragioni ma ho cambiato...
This enables us to test a candidate subspace $\Hi'$ and construct
associated QL $\{D_k\}$ by employing a variation of the randomized
algorithm presented in Sec. \ref{sec:random} for general QL
stabilization.  

Technically, the key simplification that the above invariance
requirements translates into is the fact that only {\em linear}
constraints are imposed on the noise operators.  This may be seen by
writing the block decomposition of the $D_k$ with respect to
$\Hi=\Hi'\oplus\Hi'^\perp$:
\[D_k=\left[\begin{array}{cc} D_{S,k} & D_{P,k} \\ D_{Q,k} & D_{R,k}
\end{array}\right].\] 
\noindent 
In order for $\Hi'$ to be invariant, it must hold that
\cite{ticozzi-QDS}: \beq\label{eqquad}\sum_k
{D}_{S,k}^{\dag}D_{P,k}=0\eeq and $D_{Q,k}=0$ for all $k.$ In terms of
the coefficients $\beta_{jk}$ that parametrize $D_k$ as in
Eq. \eqref{param}, the above Eq. \eqref{eqquad} is clearly a set of
quadratic equations. However, if $\Hi'^\perp$ is required to be
invariant as well, then Eq. \eqref{eqquad} becomes $D_{P,k}=0$ and all
the constraints are linear.  Hence, the proof of Theorem
\ref{randomized} (see Appendix \ref{sec:randomized}) carries over to
this case upon restricting to the subspace of $\hat\beta_{jk}$ such
that $\rho_d,$ $\Hi',$ and $\Hi'^\perp$ are invariant.  Let
$\alpha_{jk}=0$ for all $j,k$.  We thus have the following:

\begin{thm}
\label{CDrandomized}
If there exists a choice of $\beta_{jk}$ that makes $\rho_d$
$\Hi'$-DQLS and $\Hi'^\perp$ invariant, then {almost any} choice
such that $\rho_d$ and $\Hi'^\perp$ are invariant makes $\rho_d$
$\Hi'$-DQLS as well.
\end{thm}

We can thus provide the following randomized conditional stabilization
algorithm:

\begin{description}
\item{Step 0:} {\em Finding a candidate $\Hi'$.} Given Lemmi
\ref{Hperp} and \ref{Hinv}, we need $\Hi' \supsetneq \Hi_d$ such that
$\Hi'\perp \Hi_w.$ The largest, obvious candidate is of course
$\Hi'=\Hi\ominus\Hi_w.$ A trial choice may otherwise be dictated by
physical considerations and available experimental capabilities: {\em
e.g.}, in situations where a conserved quantity associated to an
observable ${\cal O}$ exists, such that $\Hi_d$ and $\Hi_w$ belong to
two different eigenvalues of ${\cal O}$, 
a way to construct $\Hi'$ is to identify it with the eigenspace containing $\Hi_d.$
%%
% This can be helpful especially in connection with the stabilizer 
% formalism \cite{nielsen-chuang}.
%% LV: Francesco mi sono persa, perche' lo stab formalism e' rilevante 
%%     qui, visto che comunque il target di sicuro non e' uno stab state?
%%     In generale, questa parte non mi era affatto chiara, ho cambiato 
%%     ma forse ci ritorno

\item{Step 1:} {\em Imposing quasi-locality of $\{D_k\}$.} Pick, as
before, a product operator basis for ${\mathfrak B}(\Hi),$ say,
$\{\sigma_{i_1}\otimes\ldots\otimes\sigma_{i_n}\,|\,$$
{i_a=1,\ldots,d_a^2}\}$. Let $D_k$ be associated to a vector $\vec
d_k,$ whose components represent our free design parameters. Impose
the QL requirement by demanding that the components associated to
basis elements that are not QL vanish: let $\hat B_{k}$ be the
orthogonal projection onto the subspace generated by the basis
elements that are QL with respect to ${\cal N}_k$, and $\hat
B_{k}^\perp=I-\hat B_k.$ Write $\hat D \equiv [\,\vec d_1|\cdots|\vec
d_n\,]$ for the matrix of all the free parameters in compact form.  We
must then require, for each $k$:
\[\hat B_{k}^\perp\vec d_k =0.\]

\item{Step 2:} {\em Ensuring invariance of $\Hi_0$.} Impose the linear
constraints $D_k\ket{\psi}=0$ for all $k$. If $\hat P_0$ is the matrix
representation of the linear (super)-operator $P_0(X)= X
\ket{\psi}\bra{\psi}$ with respect to the the chosen operator basis,
this is equivalent to require, for each $k$:
\[\hat P_0\vec d_k=0.\]

\item{Step 3:} {\em Ensuring invariance of $\Hi'$ and
$\Hi'^\perp$}. We now impose a block-diagonal form for the $D_k$ with
respect to the decompositon $\Hi=\Hi'\oplus\Hi'^\perp$.  That is, upon
writing
\[D_k=\left[\begin{array}{cc} D_{S,k} & D_{P,k} \\ D_{Q,k} & D_{R,k}
\end{array}\right],\] 
we require that \cite{ticozzi-QDS}: 
\beq
\label{eqlin} 
D_{Q,k}=0, \;\;\; D_{P,k}=0, \eeq 
\noindent 
for all $k.$ Note that Eq. \eqref{eqlin} is a linear constraint, and
can be imposed by requesting: $[D_k,\Pi']=0,$ where $\Pi'$ is the
orthogonal projector associated to $\Hi'.$ In terms of the
vectorization employed in the previous steps, this can in turn be
rewritten as: \beq \hat P'\vec d_k=0, \label{eqlin2} \eeq
\noindent 
with $\hat P'$ denoting the matrix representation of the super-operator 
associated with the commutator with $\Pi'$, $\hat P'(X)= [X,\Pi']$.
%% LV: oppure [\Pi',X]... CHECK
For each $k,$ the above (three) steps translate in the following
(homogeneous) systems of linear equations:

\beq
\label{linsys1} 
\hat C \vec d_k:= \left[ \begin{array}{c}\hat B_{k}^\perp\\ \hat
P_0\\\hat P'\end{array}\right]\vec d_k=0\eeq
\noindent 
If a non-zero solution exists, then a subspace of solutions $\vec d_k$
exist, each corresponding to a QL operator $D_k$.

\item{Step 4:} {\em Forcing convergence}. 
Choose the free variables uniformly at random within a finite interval
$[-\gamma,\gamma] \subset {\mathbb R}.$
\end{description}

We thus have the following Corollary as a direct consequence of
Theorem \ref{CDrandomized}:

\begin{cor} 
\label{alg1}
Let $\rho_d$ denote the target pure state. Assume that there exist an
invariant subspace $\Hi'$ and QL $\{D_k\}$ such that $\rho_d$ is
$\Hi'$-DQLS and $\Hi'^\perp$ is invariant.  Then the above Steps 0--4
find a choice of $\Hi'$ and $\{D_k\}$ with probability one.
\end{cor}

From a practical standpoint, Proposition \ref{alg1} provides an
explicit algorithm to test whether a candidate subspace $\Hi'$ makes
the target $\rho_d$ $\Hi'$-DQLS and, if so, it also outputs a choice
of effective noise operators $\{D_k\}$.

%---------------------------------------------------------------------------

\section{Quasi-local stabilization with drift dynamics}
\label{sec:drift}

In realistic scenarios, the uncontrolled system may evolve under an 
internal QL dynamics, due to a pre-existing 
%LV: Presence of a pre-existing non mi piaceva tanto...
Hamiltonians and/or couplings with Markovian environments, 
resulting in a non-trivial, known drift generator, ${\cal L}_0
(H_0, \{ L_k^0\})\ne 0$, in the notation of Section \ref{sec:settings}.
We first discuss the simpler case where only
Hamiltonian drift is present ($L_k^0\equiv 0$ for all $k$).

\subsection{Drift Hamiltonian}
\label{drifth}

We preliminarily note that if {\em complete} QL Hamiltonian control is
available over $S$ with respect to the same locality notion of $H_0$ 
or a less restrictive one, then a control Hamiltonian $H_c=-H_0$ 
can be applied to undo the action of the drift, given that $H_0$ is
known.  In this case, the problem becomes again effectively
drift-less, and the results developed in the previous sections apply.
We thus assume here that limited (if any) Hamiltonian control is
available.

In order to establish whether $\rho_d$ 
%can be made DQLS 
%% LV: Perche' DQLS in particolare?  Deve essere un remnant da versioni 
%%       precedenti...
can be stabilized in the presence of $H_0$, we can first check whether
$H_0\ket{\Psi}=\lambda\ket{\Psi}$, in which case the state is
invariant.  If not, the first step is to see whether invariance may be
enforced by applying suitable QL dissipation.  Consider a QL
decomposition of $H_0=\sum_k H_k^0$, and decompose $H_k^0$ in matrix
blocks according to $\Hi= \Hi_d \oplus \Hi_d^\perp$, that is:
\[H_k^0 \equiv \begin{bmatrix}H_{S,k} & H_{P,k} \\ H_{P,k}^\dag & H_{R,k} 
\end{bmatrix}.\]  
For each $k$, define:
\beq \tilde{D}_k \equiv \begin{bmatrix}1 & \tilde{D}_{P,k} \\ 0 & 0
\end{bmatrix},\quad \tilde{D}_{P,k}=2iH_{P,k}.\label{eq:comp}
\eeq 
By Corollary 1 in \cite{ticozzi-QDS}, it follows that applying
$\tilde{D}_k$ as a noise operator for each neighborhood makes $\rho_d$
invariant for the global dynamics. Therefore, by recalling Eqs.
\eqref{cond1}-\eqref{cond2} and Corollary \ref{standardform}, we can
find an equivalent representation of the generator in standard form,
say ${\cal L}(H'_0, \{ D'_k\})$, with $H'_0\ket{\Psi}=0$ and $\rho_d$
invariant under ${\cal L}$.  We may then proceed as follows:

\begin{enumerate} 
\item Determine whether $\rho_d$ would be DQLS in the absence of
$H'_0$, by applying Theorem \ref{mainthm}.

\item If $\rho_d$ is DQLS for $H_0'=0$, a straightforward modification of
Theorem \ref{randomized} above (see also Corollary \ref{Drandomized}
in Appendix \ref{sec:randomized}) proves that a {\em generic} choice
of stabilizing noise operators, $\{D_k\}$, will make $\rho_d$ GAS
under the combined evolution generated by ${\cal L}(H'_0, \{ D'_k\}
\bigcup\{D_k\})$.

\item If $\rho_d$ is not DQLS for $H'_0=0$ and some
%, yet not complete 
%% LV: Secondo me e' chiaro a questo punto, e un po' pesante ripetere?
QL Hamiltonian control is available, we still invoke Theorem \ref{randomized} 
to determine whether a stabilizing QL Hamiltonian $H_c$ (and possibly 
stabilizing noise operators  $\{D_k\}$) can be found by randomizing the 
free parameters in the controlled generator.
%we can still use Theorem \ref{randomized} after having imposed 
%invariance of the target state on the control Hamiltonian parameters. 
%% LV: Lo trovo confusing perch' sembra che imporre l'invariata sia 
%%       uno step separate?
%%       Inoltre, anche se ci sono H'_0 e i D_k', non e'  a priori garantito che 
%%       possiamo avere QLS  solo randomizzando H_c, o mi sono persa  
%%       qualcosa?
If $\rho_d$ is QLS and solutions are attainable with the available controls, 
with probability one a choice will be found, such that $\rho_d$ is GAS
under the combined generator ${\cal L}(H'_0+H_c, \{ D'_k\} \bigcup\{D_k\})$.

\item If $\rho_d$ is not DQLS and no Hamiltonian control is
available, we need to check whether $H'_0$ satisfies the necessary
conditions of Proposition \ref{effectiveH} (or Corollary \ref{effectiveH2}). If 
so, we invoke Theorem \ref{randomized} to determine whether 
stabilizing noise operators  $\{D_k\}$ can be found by randomizing the 
free parameters in the controlled generator.
% once invariance has been ensured.  
If the drift Hamiltonian $H'_0$ does not not obey Proposition \ref{effectiveH}
and $H_c\equiv 0$, $\rho_d$ {\em cannot} be made GAS.

\item If $\rho_d$ cannot be made GAS, 
%DQLS and no Hamiltonian control is available,
%% LV: Mi verrebbe da dire che si arriva al passo 5... if everything else fails?
conditional stabilization may be attempted, by invoking Theorem
\ref{CDrandomized} and a suitable modification of the randomized
stabilization algorithm.  Specifically, in Step 0 it is necessary to
additionally ensure that the candidate subspace $\Hi'$ and its complement
$\Hi'^\perp$ are invariant under the drift dynamics generated by
$H'_0$.  By denoting with $\Pi'$, as before, the orthogonal projector
onto $\Hi'$, this translates into requiring the additional
compatibility requirement $[H'_0, \Pi']=0$.  If $\rho_d$ can be made
$\Hi'$-DQLS with $\Hi'$ and $\Hi'^\perp$ invariant, the algorithm
succeeds with probability one.
\end{enumerate}

\subsection{Drift Hamiltonian and dissipation}

In the most general situation, the system may be driven by a QL
Markovian drift dynamics, specified by a generator ${\cal L}_0 (H_0,
\{ L_k^0\})$ with {\em both} $H_0$ and some noise generators being
non-vanishing.  In this case, the first key property to verify is
whether the {\em necessary condition for invariance} of $\rho_d$ is
obeyed by the dissipative drift, that is, whether for each
neighborhood, we have ${\cal N}_k$,
$L_k\ket{\Psi}=\lambda_k\ket{\Psi}$.  If
$L_k\ket{\Psi}\neq\lambda_k\ket{\Psi}$ for some $k$, then $\rho_d$
{\em cannot} be made GAS by the methods we described.  While in
practice one may expect that {\em approximate} stabilization be still
meaningful and viable if the natural dissipation is sufficiently weak
with respect to the available controlled dissipation, establishing
rigorous results in this sense requires a separate analysis, which
will be addressed elsewhere (see also \cite{WolfBounds} for relevant
distance bounds).

Let us thus assume that $L_k\ket{\Psi}=\lambda_k\ket{\Psi}$ for all
$k$.  Then by using Corollary \ref{standardform}, we can write an
equivalent generator in standard form, with operators $\{ H'_0,
L'_k\}$ such that $L'_k \ket{\Psi}=0$. By the invariance requirement,
these noise operators must satisfy Lemma \ref{supp} as well.  In this
way, we have effectively mapped the problem back the one just
considered in Sec. \ref{drifth}, with only Hamiltonian drift.  In case
%complete 
%% LV: force in case specific puo'  bastare meno?
sufficient QL Hamiltonian control is available, we can remove the effect 
of $H'_0$ as above, and the problem is {\em reduced to determine whether
$\rho_d$ is DQLS, QLS, or $\Hi'$-DQLS, as already discussed.} Otherwise
we may proceed as follows:

\begin{enumerate} 

\item If the drift Hamiltonian $H'_0$ cannot be canceled by
Hamiltonian control, we need to determine whether it destabilizes the
desired state. If so, the action of $H'_0$ can be compensated in a way
similar to the one given in Eq. \eqref{eq:comp}, but, due to the
non-zero dissipative drift $\{L'_k\}$, we now need to choose
\[\tilde{D}_k \equiv \begin{bmatrix}1 & \tilde{D}_{P,k} \\ 0 & 0
\end{bmatrix},\quad
\tilde{D}_{P,k}=2 i H_{P,k}- {L'}_{S,k}^{\dag} L'_{P,k}.\] 
\noindent 
The drift dynamics induced by $H'_0, \{ L'_k\}$, plus the QL operators
$\{\tilde{D}_k\}$, can then be associated to a new generator 
%%% ${\cal L}(H_0'', \{D_k' \equiv L''_k + D''_k\})$ 
%%% FT: No, non si possono sommare i noise operators!!! Si aggiungono solo...
%%% LV: BESTIA!  - sorry!! 
%
%${\cal L}(H_0'', \{L_k''\}) \equiv {\cal L} (\{L_k\}\bigcup \{\tilde D_k\})$
%% LV: Mi pare che questo e' il generatore giusto?... 
${\cal L}(H_0'', \{L_k''\}) \equiv {\cal L} ( H_0'', \{L_k' \}\bigcup \{\tilde D_k\})$
in standard form, with
$H_0''\ket{\Psi}=0$ and $\rho_d$ invariant under ${\cal L}$.

\item If $\rho_d$ is DQLS for $H_0''=0$, then by
Corollary \ref{Drandomized}, a {\em generic} choice of stabilizing
operators, $\{D_k\}$, will suffice to make $\rho_d$ GAS under the
combined evolution generated by ${\cal L}(H_0'', \{L''_k\}\bigcup \{D_k \})$.

\item If $\rho_d$ is not DQLS for $H_0''=0$ and some
%, yet not complete 
QL Hamiltonian control is available, we can use Theorem \ref{randomized} 
to determine whether a stabilizing QL Hamiltonian $H_c$ (and possibly 
stabilizing noise operators  $\{D_k\}$) can be found by randomizing the 
free parameters in the controlled generator.
%after having imposed invariance of the target state on the control Hamiltonian parameters. 
If $\rho_d$ is QLS and solutions are attainable with the available controls, 
with probability one a choice will be found, such that $\rho_d$ is GAS
under the combined generator ${\cal L}(H''_0+H_c, \{ L''_k\} \bigcup\{D_k\})$.

\item If the state is not DQLS or $H_0''=0$ and no Hamiltonian control is available, 
we need to check whether $H''_0$ satisfies the necessary conditions of Proposition 
\ref{effectiveH} (or Corollary \ref{effectiveH2}). If so, we are left with the problem of
finding stabilizing QL noise operators to be added to the internal
ones, by randomizing the free parameters in the total controlled
generator.  If $H''_0$ has a destabilizing action on $\rho_d$, the state is {\em
not} QLS.

\item If $\rho_d$ is neither DQLS nor QLS, conditional stabilization
may be tested, in a way similar to what described for Hamiltonian
drift only.  In Step 0 of the stabilization algorithm, we now need to
to ensure that both the candidate subspace $\Hi'$ and $\Hi'^\perp$ are
invariant under the full drift dynamics generated by ${\cal L}_0$.  
If $\rho_d$ can be made $\Hi'$-DQLS with $\Hi'$ and
$\Hi'^\perp$ invariant, the algorithm succeeds with probability one.
\end{enumerate}
 
%-----------------------------------------------------------------------------
 
\section{Illustrative examples}
\label{sec:example}

Consider a quantum register consisting of $n$ qubits, and let
$\{\sigma_i\,|\, i=x,y,z\}$ denote single-qubit Pauli
matrices \cite{nielsen-chuang}, with the corresponding multi-qubit
Pauli operators given by $\sigma_i^{(a)}\equiv I \otimes \ldots
\otimes \sigma_i \otimes \ldots \otimes I\otimes I$, and $\sigma_i$
acting on the $a$-th qubit.  Since the form of possible drift
dynamics is highly system-dependent, we focus here on the drift-less
setting, as relevant in particular to trapped-ion open-system quantum
simulators \cite{barreiro} and cold atomic gases
\cite{diehlTop}.
%% 
%% LV: Nothing on drift really?  I need to think whether some comment or 
%%     added example may be worth -- Francesco what do you think?
%% FT: Well, as we say before drift acts like it fixes, or adds some parameters in the 
%% randomization. If a state is DQLS or QLS, drift either ruins it by 
%% destabilizing \rho_d, or generically actually helps!

\subsection{DQLS states}

As mentioned earlier in the text, the DQLS states include important
classes of pure multipartite entangled states, most notably, all
stabilizer and graph states, relative to the natural choice of
neighborhoods associated with connected nodes on the graph
\cite{Kraus-entangled,ticozzi-ql}.  While characterizing the class of
DQLS states for a {\em pre-determined} neighboorhood structure remains
an interesting open problem, one can show that non-trivially entangled
pure states which are provably {\em not} graph states are in DQLS.
For instance, as already noted in \cite{ticozzi-ql}, the DQLS set for
$n=4$ qubits contains the symmetric Dicke state with two excitations,
$$D_4^{(2)} \equiv ( |0011\rangle + |0101\rangle + |1001\rangle +
|0110 \rangle + |1010 \rangle + |1100 \rangle)/\sqrt{6},$$
\noindent
which may be used as an optimal quantum resources in $1 \rightarrow 3$
telecloning protocols and has been experimentally characterized using
polarization-entangled photons \cite{Geza}.
%% LV: note that as defined by Blatt, 
%% \psi_D includes W so we do not get all Dicke states for sure
Imagine that the four qubits are arranged in a line. The state
$D_4^{(2)}$ may be projected into both the three-qubit GHZ and the W
class via single-qubit measurements. While the resulting GHZ$_3$ and
W$_3$ states are graph states (hence DQLS) relative to the natural
(`star'-connected) neighboorhood choice ${\cal N}_1 = \{1,2\}$, ${\cal
N}_2 = \{2, 1, 3\}$, ${\cal N}_3 = \{3 ,2\}$, this is no longer the
case if a nearest-neighboorhood coupling constraint is imposed, in
which case ${\cal N}_1 = \{1,2\}$, ${\cal N}_2 = \{2, 3\}$.

In what follows, we explicitly address dissipative stabilization of
$n$-qubit GHZ and W states, which play a paradigmatic role in QIP 
\footnote{We remark that dissipative preparation of entangled states 
sufficiently {\em close} to GHZ and/or W
states ({\em practical stabilization} in the control-theoretic jargon),
has been analyzed in specific physical scenarios, see e.g. \cite{Almut,Chen}.}.

\subsection{Greenberger-Horne-Zeilinger cat states}
\label{GHZ}

\subsubsection{GHZ states are not DQLS} 

A representative of the GHZ class is the state $\rho_{\text{GHZ}}
=|\Psi \rangle\langle \Psi |$, where
\beqan&&\ket{\Psi}\equiv \ket{\Psi_{\text{GHZ}}}\\&&
= (\ket{0}^{\otimes n}+ \ket{1}^{\otimes n})/\sqrt{2}\equiv
(\ket{000\ldots 0}+\ket{111\ldots 1})/\sqrt{2}.\eeqan
\noindent 
As shown in \cite{ticozzi-ql}, GHZ$_n$ states are generally not DQLS,
except in cases where the QL constraint becomes effectively trivial,
as illustrated in the above three-qubit example. In fact, any reduced
state on any (nontrivial) neighborhood is an equiprobable mixture of
$\ket{000\ldots 0}$ and $\ket{111\ldots 1}.$ It is then immediate to
see that
\[\Hi_0=\textrm{span}\{\ket{000\ldots 0}, 
\ket{111\ldots 1}\}=\bigcap_k\supp(\rho_{{\cal N}_k}\otimes
I_{\bar{\cal N}_k}), \]
\noindent
with $\text{dim}(\Hi_0) = d_0=2.$ From the point of view of MPS theory, the GHZ state is known to be non-injective, 
and corresponding to the two-fold degenerate ground state of a frustration-free 
(gapped) Ising parent Hamiltonian \cite{Norbert}.
%% DOUBLE-CHECK

\subsubsection{GHZ states can be QLS: conditions on the neighborhood size}

Let us explore under what conditions adding Hamiltonian control can
render $\rho_{\text{GHZ}}$ GAS. Given Corollary \ref{effectiveH2}, we
seek a QL Hamiltonian $H_c =\sum_k H_k $ such that: \beqan && H_c
\ket{\Psi_{\text{GZH}}}= 0,\\ && H_c \ket{\Phi_1} \notin \Hi_0, \;\;
\ket{\Phi_1} \equiv (\ket{000\ldots 0}-\ket{111\ldots 1})/\sqrt{2}.
\eeqan
\noindent 
In order for this to happen, $H_c$ must equivalently obey the
condition
\[H_c \ket{000\ldots 0}=- H_c\ket{111\ldots 1}.\]
Since each component $H_k$ is QL, it acts non-trivially on {\em at
most} a number $n_k$ of ``symbols'' (that is, 0 or 1) in the
factorized states $\ket{0}^{\otimes n},\ket{1}^{\otimes n}$,
respectively.  Let $n$ be even and assume that $n_k<n/2$.  Then it
follows that
\begin{eqnarray*}
H_c \ket{000\ldots 0} & \hspace*{-1mm}\in\hspace*{-1mm} & \textrm{span}
\Big\{\ket{x_1,\ldots,x_n}, x_j\in\{0,1\},\hspace*{-0.4mm} \sum_j x_j
\hspace*{-0.5mm}< n/2 \Big\},\\
H_c \ket{111\ldots 1} & \hspace*{-1mm}\in \hspace*{-1mm}& \textrm{span}
\Big \{\ket{x_1,\ldots,x_n}, x_j\in\{0,1\},\hspace*{-0.4mm}\sum_j x_j
\hspace*{-0.5mm}> n/2 \Big\}. 
\end{eqnarray*}
Thus, the two vectors must be orthogonal, since they belong to
subspaces spanned by two orthogonal sets of vectors. This means that a
Hamiltonian satisfying the requirements of Corollary \ref{effectiveH2}
does not exists. In other words, we need $H_c$ to be able to flip {\em
at least $n/2$ qubits} in the product basis states $\ket{0}^{\otimes
n},\ket{1}^{\otimes n}$.

If we allow one neighborhood to include $n_k=n/2$ qubits, we can
always construct a QL Hamiltonian such that \beqan && H_c
(\ket{000\ldots 0}=(\ket{1\ldots 10\ldots 0}-\ket{0\ldots 01\ldots
1})/\sqrt{2}, \\ && H_c (\ket{111\ldots 1}=(-\ket{1\ldots 10\ldots
0}+\ket{0\ldots 01\ldots 1})/\sqrt{2}, \eeqan
\noindent 
with the vectors in the r.h.s. containing {\em precisely} $n/2$ zeroes
and $n/2$ ones, which clearly satisfies the requirement.  This may be
achieved by considering a neighborhood structure consisting of two
disjoint sets ${\cal S}_{\ell=1,2}$, each including half of the qubits
if $n$ is even [or $(n+1)/2,(n-1)/2$ in the odd-$n$ case], and QL
Hamiltonian components given by
$$H_\ell=(-1)^{\ell-1}\Big(\prod_{a\in{\cal
S}_\ell}\sigma_x^{(a)}\Big).$$ 
\noindent 
If there exists two neighborhoods ${\cal N}_{k_\ell},$ compatible with
the QL constraints, such that ${\cal S}_{\ell}\subset{\cal
N}_{k_{\ell}},$ then, with a proper choice of the noise operators, a
GHZ state may be rendered GAS.  Otherwise, the above argument
proves the following:

\begin{Prop} 
Assume that no neighborhood exists, that contains $n/2$ qubits if $n$
is even, or $(n+1)/2$ if $n$ is odd. Then $\rho_{\text{GHZ}}$ is not
QLS.
\label{nogo}
\end{Prop}

While the randomized QLS algorithm may be applied if a large
neighborhood exists, stabilizing generators may be constructed by
inspection in simple low-dimensional situations.  Let us reconsider,
for example, $n=3$ qubits on a line, under the QL constraint of
two-body nearest-neighbor couplings.
% ${\cal N}_1=\{1,2\}$, ${\cal N}_2=\{2,3\}$.  
We may then choose the two disjoint sets ${\cal S}_1=\{1\},{\cal
S}_2=\{2,3\},$ and implement
\[ H_c = \sigma_x^{(1)}-\sigma_x^{(2)}\otimes\sigma_x^{(3)},\]
\[ D_1=I \otimes(\ket{00}\bra{01}+ \ket{11}\bra{10}),\]
\[ D_2=I\otimes(\ket{00}\bra{01}+ i\ket{11}\bra{10}).\]
\noindent 
We stress that the phase factor appearing in the Lindblad operator
$D_2$ is {\em not} coincidental: in fact, the more symmetric choice
$D'_2=I \otimes(\ket{00}\bra{01}+ \ket{11}\bra{10})$ would leave the
$-1$-eigenspace of the operator $\bigotimes_{a=1,2,3}\sigma_x^{(a)}$
invariant for the Liouvillian generator associated to $H_c, D_1,
D'_2$.  In constrast, $\rho_{\text{GHZ}}$ is the {\em unique}
invariant state for the generator associated to $H_c, D_1, D_2$, as
required for GAS.

This simple example is sufficient to establish that DQLS $\subsetneq$
QLS, namely that there exist pure entangled states that are not
stabilizable by dissipation alone but can indeed be made GAS by the
addition of a suitable QL Hamiltonian.  Furthermore, our argument for
rendering GHZ states GAS with QL resources is general, and, in the
light of Proposition \ref{nogo}, it requires neighborhoods of the
minimum possible size.  Since the latter scales linearly with the
number of qubits $n$, the order of interaction in the required control
Hamiltonians likewise grows extensively with $n$, preventing QLS to be
achievable in a scalable fashion.

\subsubsection{GHZ states are conditionally DQLS with a scalable strategy}

Consider again a three-qubit register, with locality notion associated
to ${\cal N}_{1}=\{1,2\}$, and ${\cal N}_{2}=\{2,3\}$.  In order to
show that GHZ states can be $\Hi'$-DQLS for a properly chosen $\Hi',$
note that $\ket{\Psi}_{\text{GHZ}}$ is a $+1$-eigenvector of
$\sigma_x^{\otimes 3}$, whilst
$\ket{\Phi_1}=(\ket{000}-\ket{111})/\sqrt{2}$ is a $-1$-eigenvector of
the same observable. In the notation of Section \ref{sec:conditional},
we have $\Hi_w=\span\{ \ket{\Phi_1} \}.$ We can thus try to choose
$\Hi'$ as the $+1$-eigenspace of $\sigma_x^{\otimes 3},$ denoted by
$\Hi_{xxx}^+$.  As noise operators, pick $D_{12}=D\otimes I,$ and
$D_{23}=I\otimes D,$ with \beq D=\left[\begin{array}{cccc} 0 &0 &1 &0
\\ 0 &0 &0 &0 \\ 0 &0 &0 &0 \\ 0 &1 &0 &0\end{array}\right] \equiv
\sigma_+ \otimes \frac{\sigma_z+I}{2} - \sigma_- \otimes
\frac{\sigma_z -I}{2},
\label{D}
\eeq
%% CHECK that above is right...
\noindent 
in the standard Pauli formalism.  It is easy to show that they both
commute with $\sigma_x^{\otimes 3},$ and leave $\rho_{\text{GHZ}}$
invariant. Finally, by running the DID algorithm for GAS verification
described in Appendix \ref{algorithm} starting from
$\Hi_0=\textrm{span}\{\ket{000}, \ket{111}\}$ as the target subspace,
one finds that it runs to completion in 4 steps and hence $\Hi_0$ is
GAS.  Hence, by Theorem \ref{conditionalstab}, we conclude that
$\rho_{\text{GHZ}}$ is $\Hi_{xxx}^+$-DQLS.

Does the same strategy still work for a generic number of qubits?
More precisely, consider $n$ qubits arranged on a linear graph
(equivalently, an open spin chain), with neighborhoods associated to
all nearest-neighbor pairs, namely ${\cal N}_k=\{k,k+1\}_{k=1,\ldots,
n-1}.$ Implement a dissipator for each neighborhood of the form
$D_k=D\otimes I_{\bar{\cal N}_k},$ where $D$ has the structure given
in \eqref{D}. Consider, as in the above three-qubit case, any initial
state with support on $\Hi',$ the $+1$-eigenspace of the operator
$\sigma_x^{\otimes n}.$ Explicitly checking GAS of the $n$-qubit
subspace $\Hi_0$ by resorting to the DID algorithm becomes impractical
as $n$ grows. Luckily, Theorem \ref{conditionalstab} makes this step
unnecessary.  In fact, note that:
\begin{enumerate}
\item This choice of $D_k$ makes ${\mathfrak D}(\Hi^\circ_{{\cal
N}_k})$ GAS on {\em any} neighborhood, so that ${\mathfrak D}(\Hi_0)$
is GAS;
\item The $D_k$ commute with $\sigma_x^{\otimes n}$ and leave $\Hi'$
invariant;
\item The other state in $\Hi_0,$ $\ket{\Phi_1}$ does not belong to
$\Hi'.$
\end{enumerate}
We can thus directly conclude that the GHZ$_n$ state is $\Hi'$-DQLS,
with the same choice of two-body dissipators for any $n$ and the
dimension of the conditional preparation subspace being equal to
$d/2=2^{n-1}$.  In practice, initialization to any state in $\Hi'$ by
either coherent or incoherent means will suffice.  For instance, two
possible {\em scalable} strategies are initialization into the product
state $\ket{+}^{\otimes n}$ by application of a collective Hadamard
gate $H^{\otimes n}$ to $\ket{0}^{\otimes n}$ (in case the latter is a
natural starting point) or by a projective measurement of the
collective (one-body) spin observable $S_x\equiv \sum_a
\sigma_x^{(a)}$, post-selected on the outcome corresponding to the
maximal eigenvalue (highest-weight) state.

\subsection{W states}

\subsubsection{W states are not DQLS} 

A representative of the W class is the state $\rho_{\text{W}} =|\Psi
\rangle\langle \Psi|$, with
$$\ket{\Psi}\equiv \ket{\Psi_{\text{W}}}=(\ket{100\ldots
0}+\ket{010\ldots 0}+\ldots +\ket{000\ldots 1})/\sqrt{n}.$$
\noindent 
This state is a permutation-invariant superposition of all
computational basis states with a single 1, thus also a symmetric
Dicke state one-excitation.  The reduced states on any non-trivial
neighborhood are statistical mixtures of $\ket{000\ldots 0}$ and a
smaller W state, say $\ket{\Psi_{\text{W}'}}$, whose dimension is
determined by the neighborhood.  Accordingly, as established in
\cite{ticozzi-ql}, $\rho_{\text{W}}$ is generally not DQLS, since
\[\Hi_0=\textrm{span}\{\ket{000\ldots 0}, \ket{\Psi_{\text{W}}} \}
=\bigcap_k\supp(\rho_{{\cal N}_k}\otimes I_{\bar{\cal N}_k}),\] 
with $\text{dim}(\Hi_0) = d_0=2.$ Like the GHZ state, the W state is also known to be non-injective, 
however a frustration-free gapped parent Hamiltonian does not 
exist in this case \cite{Norbert}. 

\subsubsection{W states can be QLS: two-body interactions}

We here show that W states can be made QLS in the presence of QL
constraints that prevent DQLS to be achievable.  Consider, in
particular, a $n$-qubit register with neighborhoods associated to
arbitrary subsystem pairs, that is, $ {\cal N}_{jk}=\{j,k\}$,
$j,k=1,\ldots, n$, $j \ne k$.  Since
$\Hi_0=\span\{\ket{\Psi_W},\ket{0^{\otimes n}}\},$ we are in a
situation where Corollary \ref{effectiveH2} applies.

In order to construct a control Hamiltonian that satisfies the
requirements of Corollary \ref{effectiveH2}, it is useful to note that
the target state may be rewritten as
\beqan \ket{\Psi_{\text{W}}} = &&\sqrt{\frac{n-2}{n}} \,\ket{0}
\ket{\Psi_{\text{W}_{n-2}}} \ket{0}\\&& + \frac{1}{\sqrt{n}} \,(\ket{1}
\ket{0}^{\otimes (n-2)}\ket{0} \\&&+ \ket{0} \ket{0}^{\otimes (n-2)}
\ket{1}). \eeqan
\noindent 
It is then easy to show that, for example, the following two-body
Hamiltonian satisfies the desired conditions:
$$ H_c = \sigma_x^{(1)} P_0 - P_0 \sigma_x^{(n)}, \;\;\; P_0 =
\sum_{a=2}^{n-1}\sigma_z^{(a)}-(n-4)I^{(a)}.$$ 
\noindent 
This follows from the fact that $P_0$ is an operator on
$\bigotimes_{a=2}^{n-1}\Hi_a$ which obeys  
\[P_0\ket{\Psi_{W_{n-2}}}=0,\quad P_0\ket{0}^{\otimes (n-2)}=
\ket{0}^{\otimes (n-2)}.\] 
\noindent 
While we derived this particular Hamiltonian guided by simple symmetry
considerations, to our scope it suffices to verify by direct
computation that
\[H_c \ket{\Psi_{W}}=0,\quad H_c \ket{0}^{\otimes n}=
\ket{10\ldots 0}-\ket{0 \ldots 01},\]
\noindent 
as needed for Corollary \ref{effectiveH2}.  In order to conclude that
the above Hamiltonian indeed makes $|\Psi_W\rangle$ QLS, we also need
to exhibit an explicit choice of dissipators $D_k$.  To this aim,
consider in each two-qubit neighborhood a ladder dissipator of the
form
$$D  =\left[\begin{array}{cccc} 
0 &1 &1 &0 \\ 0 &0 &0 &1 \\ 0 &0 &0 & 1\\ 0 &0&0&0 
\end{array}\right] \equiv I \otimes \sigma_+ + \sigma_+ \otimes I. $$ 
%\frac{1}{\sqrt{2}}\ket{00}(\bra{01}-\bra{10}) +
%\frac{1}{\sqrt{2}}(\ket{01}+\ket{10})\bra{11}\}. 
%% CHECK 
\noindent 
This choice makes the set of states with support on $\Hi^\circ_{{\cal
N}_k}$ GAS on each neighborhood, and hence $\Hi_0$ is stabilized.

Explicit numerical calculation demonstrates that the above combination
of Hamiltonian and dissipative control makes $|\Psi_W\rangle$ GAS, and
hence QLS, at least for $n=3,4$.  While we have no general formal
proof, we expect that the same protocol will work for arbitrary $n$.
This shows how W states can be made QLS by allowing for {\em
arbitrary} two-body interactions.  Whether scalable stabilization
protocols may be constructed in other relevant QL scenarios, for
instance solely involving nearest-neighbors two-body interactions,
remains open to further investigation.

\subsubsection{W states are conditionally DQLS with a scalable strategy}
 
Interestingly, W states can be conditionally stabilized by employing
the systematic approach of Section \ref{wtype}.

The key step is to show that the strict inclusion condition given in
Eq. \eqref{nonover} holds. As we already noticed, with respect to
any neighboorhood topology we have $\Hi_0=\text{span}
\{\ket{\Psi_W},\ket{0}^{\otimes n}.\}$ 
%Define $d_k=\#({\cal N}_k),$
%% ?  LV: E' questo che volevi dire?
%% FT: SI!
Let $d_k$ denote the number of subsystems in the neighboorhood ${\cal
N}_k$ and, as before, let $\ket{\Psi_{W_{d_k}}}$ be a W state on $d_k$
qubits. We thus have to check, if for every $k$:
\beqan \text{supp}\left( \tr_{{\bar{\cal N}_k}}(\ket{0}\bra{0}^{\otimes n})\right)
&=&\text{span}\{\ket{0}^{\otimes d_k}\}\\&=&\Hi^w_{{\cal N}_k}\subsetneq
\Hi^\circ_{{\cal
N}_k}\\&=&\text{span}\{\ket{\Psi_{W_{d_k}}},\ket{0}^{\otimes d_k}\},\eeqan
\noindent 
which is clearly true.  Consider then
\beqan\Hi'=\Hi\ominus\bigcap_k\tilde\Hi_k&=&\Hi\ominus\bigcap_k
\text{span}\{\ket{0}^{\otimes d_k}\}\otimes \Hi_{\bar{\cal
N}_k}\\&=&\Hi\ominus \text{span}\{\ket{0}^{\otimes n}\}.\eeqan
\noindent 
Physically, we may think of $\Hi'$ as the subspace of states
orthogonal to the vacuum, thus initialization in $\Hi'$ may be
achieved in principle by any (coherent or incoherent) means that
creates at least one ``excitation''.

In each neighborhood, consider now $\ket{0}^{\otimes
d_k},\ket{\Psi_W^{d_k}}$ in this order and complete it to an
orthonormal basis for $\Hi_{{\cal N}_k}.$ With respect to this basis,
define the following ladder-type Lindblad operator:
\[D_{{\cal N}_k}=\left[\begin{array}{c|c|ccc} 0 & 0 & &\cdots 
&\\\hline 0 & 0 & 1 & 0 & \cdots\\ \hline \vdots & 0 & 0 & 1 &
\ddots\\ & & & \ddots & \ddots\\
\end{array}\right].\]
The corresponding dissipative process may be thought as cooling the
system to the reduced W state $\ket{\Psi_W^{d_k}}$, while leaving the
(QL) ground state $\ket{0}^{\otimes d_k}$ invariant.  Then construct
the overall dissipators as $D_k=D_{{\cal N}_k}\otimes I_{\bar{\cal
N}_k}.$ This choice ensures that:
\begin{enumerate}
\item $\Hi'$ is invariant for each $D_k$;
\item ${\mathfrak D}(\Hi_0)$ is GAS.
\end{enumerate}
Corollary \ref{corwtype} may then be invoked to establish that
$\ket{\Psi_W}$ is $\Hi'$-DQLS for the choice of $\Hi'$ and $\{D_k\}$
we made. Remarkably, the proposed protocol is both {\em scalable and
portable}, in the sense it works for an arbitrary number of qubits and
arbitrary QL notions.

%--------------------------------------------------------------------------------------------

\section{Conclusion and Outlook}
\label{sec:end}

We have provided a system-theoretic analysis of different 
scenarios and strategies for designing Markovian evolutions that have a desired 
pure entangled state as their unique stable steady state, using 
time-independent control parameters and subject to realistic locality 
constraints.  In particular, we have shown how target states that are 
not stabilizable under purely dissipative QL control, as previously 
considered in \cite{ticozzi-ql}, may be dissipatively prepared
upon restricting the allowed set of initializations or by allowing a 
combination of Hamiltonian and dissipative control. We have further 
addressed the role of Hamiltonian and/or dissipative drift dynamics, as 
arising from possible always-on coherent interactions and/or couplings 
of the target system to an uncontrollable Markovian environment.  
Constructive algorithms for synthesizing effective choices of stabilizing 
Markovian generators have been presented, suitable in principle for 
open-loop control implementations based on a switching output-feedback 
law along the lines described in \cite{ticozzi-ql}. In particular, if the Markovian 
semigroup is obtained as an average over the trajectories of a stochastic 
master equation, we recall that convergence of the semigroup entails 
convergence of the underlying stochastic dynamics in probability 
\cite{ticozzi-stochastic}.

While our results substantially expand the theoretical framework and 
toolbox for QL dissipative entanglement engineering in QIP, a number of 
open questions and further directions for exploration exist.  
For time-independent Markovian dynamics as considered thus far, issues of 
{\em efficiency} and {\em robustness} are especially important from a practical 
standpoint: (i) On the one hand, once a target state of interest is found to 
be stabilizable, it is desirable to characterize the {\em speed} with which asymptotic 
convergence is attained depending on the system size and the elapsed 
(in reality always finite) stabilization time.  While the  analysis carried out in 
\cite{Verstraete2009} indicates that the relevant Liouvillian spectral gap scales 
favorably with $n$ at least for injective MPSs, additional work is needed for 
more general classes of states as well as for optimally ``tuning'' the 
convergence speed as a function or the available control parameters, 
in the spirit of \cite{ticozzi-NV}.  (ii) On the other hand, assessing how 
{\em sensitive} steady states are with respect to deviations of the actual 
control parameters with respect to the intended ones, and/or to additional 
(static or time-dependent, possibly quantitatively unspecified) perturbations 
is worth being pursued both in terms of analytical bounds and numerical 
exploration in specific QIP settings. General results recently established in 
\cite{WolfBounds,Rouchon} may prove useful in that respect.

Partly related to the above, given that finite evolution time and errors will 
typically prevent a pure steady state to be exactly achieved, addressing the 
general problem of dissipative QL mixed-state stabilization is also a natural 
important next step.   Additional physical motivation to consider target mixed 
entangled states is provided by the possibility to characterize and engineer 
non-equilibrium critical behavior in noise-driven many-body systems, so-called 
{\em dissipative phase transitions}, as recently investigated e.g. in 
\cite{Carmichael,Eisert,Horstmann,Schwager} and already experimentally 
explored in \cite{schindler} in small dimension. 

Finally, a yet different direction is provided by the investigation of dissipative 
entanglement engineering in different classes of open quantum dynamical models: 
from this point of view, both switched QL Lindblad dynamics generalizing on the 
work of \cite{Thomas}, and discrete-time Kraus-map engineering \cite{Baggio} 
are especially promising in the near term.  Developing a stabilization framework for 
continuous-time non-Markovian dynamics is also an important challenge down the 
line, as relevant to both understanding the role of non-Markovianity as a resource 
\cite{Huelga} and to open-system quantum simulators \cite{barreiro,Mataloni,Sarah}.

\section*{Acknowledgements}
\noindent
L.V. is grateful to Gerardo Ortiz for discussions and Norbert Schuch for 
input on MPS theory. F.T. and L.V. acknowledge support by the QUINTET and the QFuture 
projects of the University of Padova, Italy.  L.V. acknowledges partial support from the 
NSF through grant number PHY-1104403 and hospitality and partial support 
from the \emph{Kavli Institute of Theoretical Physics} at UCSB, 
where part of this work was completed.

%------------------------------------------------------------------------------

%\bibliographystyle{ifacconf}
%\bibliography{bib-11-2012}
%\bibliography{ifacconf}             
                                     
%\section*{References}                                                   

%%%%%%%%%%%%%%%%%%%%%%%%%%%%%%%%%%%%%%%%%%%%%%%%%%%%%%%%%%%%%%%%

\appendix{: Linear-algebraic results}
%% LV: good title?

We establish here a preliminary result that will be instrumental in
the proof of the main Theorem \ref{randomized}.  Let us first recall a basic result from the theory of analytic functions:
\begin{lemma} \label{polsol} 
Let $f(x):\R^K\rightarrow \R^M$ be a (non-zero) analytic function in
$x\in\R^K$ and let ${\cal S} = \{x\in\R^K\,|\,f(x) = 0\}.$ Then
$\mu({\cal S})=0,$ where $\mu$ is the Lebesgue measure in $\R^K$.
\end{lemma}

Define an $m\times n$ matrix $X=[f_{jk}(x)],$ with
$f_{jk}:\R^K\rightarrow\C$, such that the real and imaginary parts
$\Re(f_{jk}),\Im(f_{jk})$ are (real)-analytic, and let $r_m \equiv
\max_{x\in\C^K}\text{rank}(X).$ Notice that
$\text{rank}(X)\in\{0,\ldots,\text{min}\{n,m\}\}$ for all $x\in \C^K,$ and
the maximum is attained in $\R^K$.   We thus have the following:

\begin{lemma}
\label{solmeasure} 
The set ${\cal X}=\{x\in \R^K\,|\,\rank(X)<r_m\}$ is such that
$\mu({\cal X})=0.$ 
\end{lemma}

\vspace*{12pt}
\noindent
{\bf Proof:} Let $\hat x\in\R^K$ be such that $\hat X=[f_{jk}(\hat
x)]$ has $\text{rank}(\hat X)=r_m.$ Then there exists a square
$r_m\times r_m$ submatrix $X_{m}$ of $\hat X$ that has rank $r_m$
\cite{horn-johnson}. Notice that $\det(X_m)$ is a (complex) polynomial
function of the elements of $X,$ $f_{jk},$ and hence of
$\Re(f_{jk}),\Im(f_{jk}).$ Thus, $\Re(\det(X_m))$ and $\Im(\det(X_m))$
are (real)-analytic functions of $x\in\R^n,$ and by virtue of the
previous Lemma the set: 
\beqan{\cal X}&=&\{x\in
\R^K\,|\,\det(X_m)=0\}\equiv \{x\in
\R^K\,|\,\Re(\det(X_m))\\&=&0=\Im(\det(X_m))\}\\&=&\{x\in
\R^K\,|\,\text{rank}(X)<r_m\}\eeqan
\noindent 
is the intersection of two zero-measure set.  Thus, it itself
satisfies $\mu({\cal X})=0.$ \qed

\appendix{: Dissipation-induced decomposition and randomized
stabilization}
\label{sec:randomized}

A general approach to decide the stability of a subspace, or more
precisely, of the set of states with support on a subspace, has been
developed in \cite{ticozzi-NV}. We recall here some basic ideas and
results, expanding the (partially randomized) strategy for Hamiltonian
control design proposed there to the synthesis of both coherent and
dissipative controls.

Let $\Hi_S$ be a proper subspace of $\Hi$ and ${\mathfrak
D}(\Hi_S)\subsetneq {\mathfrak D}(\Hi)$ the set of states with support
on $\Hi_S.$ Then ${\mathfrak D}(\Hi_S)$ is GAS for the QDS dynamics if
and only if a Hilbert space decomposition in orthogonal subspaces, of
the form \beq
\label{DID}
\Hi=\Hi_S\oplus\Hi^{(1)}_T\oplus\Hi^{(2)}_T\ldots\oplus\Hi^{(q)}_T,\eeq
can be obtained as the output of a constructive algorithm for GAS
verification \cite{ticozzi-NV} (see also next section). Such
decomposition is called the {\em Dissipation-Induced Decomposition}
(DID).  Each of the subspaces $\Hi^{(i)}_T$ in the direct sum is
referred to as a {\em basin}.

Partitioning each matrix associated to the noise operators $D_k$ in
blocks according to the DID results in the following standard
structure, where the upper block-diagonal blocks establish the
dissipation-induced, cascade connections between the different basins
$\Hi_T^{(i)}:$
\begin{eqnarray}
D_k=
\left[\begin{array}{c|cccc}  
D_S & \hat D_P^{(0)} & 0 & \cdots &  \\
\hline	0 &  D_T^{(1)} & \hat D_P^{(1)} & 0 & \cdots\\
\vdots & D_Q^{(1)} & D_T^{(2)} & \hat D_P^{(2)} & \ddots\\
 &  \vdots & \ddots & \ddots & \ddots\\
\end{array}\right]_k .
\label{matDID}
\end{eqnarray}
Similarly, for the control Hamiltonian we get:
\begin{eqnarray}
H_c =
\left[\begin{array}{c|cccc}  
H_S &  H_P^{(0)} & 0 & \cdots &  \\
\hline	H_P^{(0)\dag} &  H_T^{(1)} & \cdots &  & \\
0 & \vdots & \ddots &  & \\
\vdots &   &  &  & \
\end{array}\right]_k . 
\label{matHDID}
\end{eqnarray}
By construction, the $\hat D_P^{(i)}$ blocks are either zero or full
rank. The fact that the first column of blocks has only $D_S\neq 0$ is
a necessary condition for the invariance of ${\mathfrak D}(\Hi_S).$ It
follows that $\hat D_P^{(0)}\neq 0,$ otherwise ${\mathfrak D}(\Hi_S)$
cannot be GAS.

Consider now the target pure state $\rho_d=\ket{\Psi}\bra{\Psi},$ its
support $\Hi_d \equiv \Hi_S,$ $\Hi_d^\perp \equiv \Hi_R,$ and
$\Hi=\Hi_S\oplus\Hi_R$. Assume that we are free to design $H,\{D_k\}$
in parametric form, as given in Eq. \eqref{param}.
%\[H=\sum_j\alpha_{jk}\sigma_{jk},\quad D_k=\sum_j\beta_{jk}\sigma_{jk},\]
%where $\alpha_{jk},\beta_{jk}$ are chosen at random with uniform
%distribution in an interval $[-\gamma,\gamma]$ of the real axis, and
%$\{\sigma_{jk}\}$ is an operator basis (with respect to the real
%field, i.e. doubling the dimension w.r.t. the more usual complex
%field) for QL operators on ${\cal N}_k$.  
The DID provides a key tool for proving Theorem \ref{randomized},
which we reproduce below:

{\bf Theorem 2} {\em If there exists a choice of
$\alpha_{jk},\beta_{jk} \in {\cal I}=[-\gamma,\gamma]$ that makes $\rho_d$ QLS, then almost
any choice of $\alpha_{jk},\beta_{jk}\in {\cal I}$ that makes $\rho_d$
invariant, makes it QLS as well.}
%% LV: Francesco, la proof mi sembra a posto - ma e' un po' dura da 
%%     seguire.  A parte questo, dubito che a me sarebbe venuta in mente!
%% FT: Contorsioni... e` la mia specialita` a "yoga" ;)
%% LV: Dai!  Ti sei convinto a fare yoga??  Ti vedo bene a dare upward facing dog... ;)

\vspace*{12pt}
\noindent
{\bf Proof:} We begin by observing that if a choice of controls
$H_c,\{D_k\}$ make ${\mathfrak D}(\Hi_S)$ GAS, then for any
$\lambda\in\R$ the rescaled controls $\lambda H_c
,\{\sqrt{\lambda}D_k\}$ also make it GAS, since the total Liouvillian
generator is linear in $H_c$ and quadratic in $D_k.$ Assume then that
a certain choice of $\alpha_{jk},\beta_{jk}$ makes ${\mathfrak
D}(\Hi_S)$ invariant: one can always rescale all
$\alpha_{jk},\beta_{jk}$ by $\lambda$ small enough so that they are
all in $[-\gamma,\gamma]$ and still obtain a viable solution. Hence,
it is not restrictive to look for stabilizing parameters in a bounded
interval.

Next, assume that $\alpha_{jk},\beta_{jk}$ make ${\mathfrak D}(\Hi_S)$
invariant. Since, as shown in Section \ref{standardf}, it is not
restrictive to assume that $H_c,\{D_k\}$ are in standard form, we may
take $\Hi_S\subseteq\text{ker}(D_k)$ for all $k.$
The invariance conditions then translate in linear constraints
on $\alpha_{jk},\beta_{jk},$ namely:
\beq\sum_j\alpha_{jk}\sigma_{jk}\ket{\Psi}=0,\quad
\sum_j\beta_{jk}\sigma_{jk}\ket{\Psi}=0,
\label{invarianceconds}\eeq
Since by hypothesis $\rho_d$ is GAS, there must exist at least a
solution.  If the solution is unique, the statement is trivial. If the
solution is not unique, by linearity there is a subspace of
solutions. If so, we may re-parametrize the admissible solutions in
the free-parameters $\hat\alpha_{jk},\hat\beta_{jk}\in \R.$
 
We now focus on a DID that makes $\rho_d$ GAS, and that will be proved
to be generic. By construction, the matrix block decomposition of the matrix
representation of $H_c,\{D_k\}$ must be such that for each iteration,
indexed by $j,$ either \beq\tilde{\cal D}^{(j)}:=
\left[\begin{array}{c}D_{P,1}^{(j)}\\\vdots \\
D_{P,M}^{(j)}\end{array}\right]\label{rp1}\eeq 
\noindent 
has {\em maximum rank}
$r^{(j)}=\max\{\dim(\Hi_T^{(j-1)}),\dim(\Hi_T^{(j)})\}$ (see Step 2, 
3.a and 3.b), or \beq\tilde{\cal L}^{(j)}_P:=i
{H}_P^{(j)}-\frac{1}{2}\sum_k
{D}_{Q,k}^{(j)\dag}D_{T,k}^{(j)}\label{rp2}\eeq
\noindent 
has {\em full rank} $r^{(j)}$ (defined as above, see step 3.c of the
algorithm).  At the $j$-th iteration, $\dim(\Hi_T^{(j-1)})$ is fixed,
but we can choose a set of parameters
$\hat\alpha_{jk},\hat\beta_{jk}$ that maximizes $\dim(\Hi_T^{(j)}).$
Given Eq. \eqref{rp1}-\eqref{rp2}, this is equivalent to maximize the
rank of either $\tilde{\cal D}^{(j)}$ or $\tilde{\cal L}^{(j)}_P$ at
each iteration.  The elements of $\tilde{\cal D}^{(j)}$ and
$\tilde{\cal L}^{(j)}_P$ are (complex) polynomial functions of the
real parameters $\hat\alpha_{jk},\hat\beta_{jk},$ and hence Lemma
\ref{solmeasure} ensures that the parameter set corresponding to
maximal rank of the corresponding matrices has measure $1.$
%\footnote{Notice this is true since we used real parameters, 
%otherwise the adjoint in the definition of $\tilde{\cal L}^{(j)}_P$ 
%would introduce the complex conjugation of the parameters, which 
%is not in general analytic.} 
This implies, in turn, that the {\em same DID is constructed with
probability one when the parameters are randomly chosen} as above,
establishing that the target can be made GAS 
%by randomizing the parameters 
with probability one, as claimed.  \qed

Along similar lines, we may show that the set of parameters that makes 
$\rho_d$ invariant has measure zero. 
In fact, such parameter choices correspond to the proper linear hyperplane 
% hyperspace 
%%  LV: hyperspace = http://en.wikipedia.org/wiki/Hyperspace_%28topology%29
%%  E' questo che intendi??  Ho messo hyperplane per ora....
determined by Eq. $\eqref{invarianceconds}.$ The
same holds for the DQLS property: the proof is identical to the one
above, just consider the $\alpha_{jk}$ fixed. Thus, if
there exists a choice of $\beta_{jk}$ that makes $\rho_d$ DQLS, then
almost any choice of $\beta_{jk}$, such that $\rho_d$ is invariant,
makes it DQLS as well.

\begin{cor}
\label{Drandomized}
Assume $H_c =\sum_k H_k$ to be fixed, that is, $\alpha_{jk}$ are given
for all $j,k.$ If there exists a choice of $\beta_{jk} \in I$ that
makes $\rho_d$ DQLS, then almost any choice of $\beta_{jk} \in I$,
such that $\rho_d$ is invariant, makes it DQLS as well.
\end{cor}

As discussed in the main text, the randomized approach carries over to
conditional stabilization also, provided that care is taken in
ensuring that no quadratic constraints appear. While imposing
invariance of both $\Hi'$ and its complement $\Hi'^\perp$ is a
mathematically natural restriction to circumvent this issue,
it need not always be easy to ensure in physical systems, 
especially in the presence of drift dynamics.  Thus, an interesting
open question is to determine whether a randomized approach may be
devised for conditional DQLS in full generality.

\appendix{: DID algorithm}
\label{algorithm}

In order to make our presentation self-contained, we reproduce here the
algorithm for the construction of the DID \cite{ticozzi-NV}.  The
initial (target) subspace $\Hi_d\equiv \Hi_S$ (or, better, ${\mathfrak
D}(\Hi_S)),$ is GAS if and only if the algorithm runs to completion.

The inputs are a QDS generator ${\cal L}$ specified by the operators
$H,\{L_k\}$ and an invariant subspace $\Hi_S$ or the evolution, that
is, such that $e^{{\cal L}t}{\mathfrak D}(\Hi_S)\subseteq {\mathfrak
D}(\Hi_S).$ Checking whether the above property holds can be easily
done using Proposition \ref{prop1}.

\noindent \rule{\linewidth}{1pt} \\[-1mm] {\em Algorithm for GAS
verification and DID construction} \\ [-2.5mm]
\noindent \rule{\linewidth}{1pt}  

Let $\Hi_S$ be invariant. Call $\Hi_R^{(0)}:=\Hi_R,$
$\Hi_S^{(0)}:=\Hi_S,$ choose an orthonormal basis for the subspaces
and write the matrices with respect to that basis. Rename the matrix
blocks as follows: $H_{S}^{(0)}:= H_{S},$ $H_{P}^{(0)}:= H_{P},$
$H_{R}^{(0)}:=H_{R},$ $L_{S,k}^{(0)}:=L_{S,k},$
$L_{P,k}^{(0)}:=L_{P,k}$, and $L_{R,k}^{(0)}:=L_{R,k}.$ 

For $j\geq 0$, consider the following iterative procedure:

\begin{enumerate}
\item Compute the matrix blocks $L_{P,k}^{(j)}$ according to the
decomposition ${\Hi}^{(j)}=\Hi_S^{(j)}\oplus\Hi_R^{(j)}.$
\item Define $ \Hi_R^{(j+1)}:=\bigcap_k\ker L_{P,k}^{(j)}.$
\item Consider the following three sub-cases:
\begin{itemize}
\item[a.] If $ \Hi_R^{(j+1)}=\{0\}$,  define $\Hi_T^{(j+1)}:=\Hi^{(j)}_R.$ 
The iterative procedure is successfully completed.

\item[b.] If $ \Hi_R^{(j+1)}\neq\{0\},$ but $ \Hi_R^{(j+1)}\subsetneq
\Hi_R^{(j)},$ define $\Hi_T^{(j+1)}$ as the orthogonal complement of
$\Hi_R^{(j+1)}$ in $\Hi_R^{(j)}$, that is,
$\Hi_R^{(j+1)}=\Hi_R^{(j)}\ominus\Hi_R^{(j+1)}.$ 

\item[c.] If  $ \Hi_R^{(j+1)}= \Hi_R^{(j)}$ (that is, $L_{P,k}^{(j)} =
0 \; \forall k$), define
$$\tilde{\cal L}_P^{(j)}:=-i {H}_P^{(j)}-\frac{1}{2}\sum_k
{L}_{Q,k}^{(j)\dag}L_{R,k}^{(j)}.$$
\begin{itemize}
\item If $\tilde{\cal L}_P^{(j)}\neq0,$ re-define
$\Hi_R^{(j+1)}:=\ker(\tilde{\cal L}^{(j)}_P)$.\\ If $
\Hi_R^{(j+1)}=\{0\}$, define $\Hi_T^{(j+1)}:=\Hi^{(j)}_R$ and the
iterative procedure is successfully completed. Otherwise define
$\Hi_T^{(j+1)}:=\Hi_R^{(j)}\ominus\Hi_R^{(j+1)}$.

\item If $\tilde{\cal L}^{(j)}_P=0,$ then $\Hi_R^{(j)}$ is invariant
and $\Hi_S$ cannot be GAS. Exit the algorithm.
\end{itemize}

\end{itemize}

\item Define $\Hi_S^{(j+1)}:=\Hi_S^{(j)}\oplus \Hi^{(j+1)}_T.$ To
construct a basis for $\Hi_S^{(j+1)},$ append to the {\em already
defined} basis for $\Hi^{(j)}_S$ an orthonormal basis for $
\Hi^{(j+1)}_T.$

\item Increment the counter $j$ and go back to step 1.
\end{enumerate}

\noindent \rule{\linewidth}{1pt} \\

\vspace*{1mm}

\noindent 
The algorithm ends in a finite number of steps, since at every
iteration it either stops or the dimension of $\Hi^{(j)}_R$ is reduced
by at least one.

\end{document}